\documentclass[sigconf, nonacm]{acmart}

\usepackage{algorithm}
\usepackage{algpseudocode}
\usepackage{enumitem}
\usepackage{graphicx}
\usepackage{subfig} % preamble에 있어야 함
\usepackage{subcaption} % subfigure 대신 최신 패키지
\usepackage{balance}

\usepackage{tikz}
\usetikzlibrary{calc}
\usepackage[most]{tcolorbox}
\usepackage[mnwidth=2cm,boxcolour=orange]{mnotes}
\usepackage{todonotes}

\newsavebox{\pipelineAbox}
\newsavebox{\pipelineBbox}
\newlength{\pipelinegap}
\newlength{\pipelineH}

\newcommand{\frameworktitle}{\textsc{SPDP}} %% Name of Our Framework 
\usepackage[a-2b]{pdfx}          % Produce PDF/A-2b
\setRGBcolorprofile{sRGB.icc}{sRGB IEC61966-2.1}{sRGB}{}

\newcommand{\rb}[1]{\textcolor{black}{#1}}

\newcommand{\setpipelineheight}[3][0pt]{%
  \setlength{\pipelinegap}{#1}%
  \sbox{\pipelineAbox}{\includegraphics[height=1pt]{#2}}%
  \sbox{\pipelineBbox}{\includegraphics[height=1pt]{#3}}%
  \setlength{\pipelineH}{\linewidth-\pipelinegap}%
  \setlength{\pipelineH}{%
    \pipelineH * \ratio{1pt}{\wd\pipelineAbox+\wd\pipelineBbox}%
  }%
}
\newcommand\vldbdoi{10.14778/3836663.3836665}
\newcommand\vldbpages{2950 - 2963}
\newcommand\vldbvolume{19}
\newcommand\vldbissue{11}
\newcommand\vldbyear{2026}
\newcommand\vldbauthors{\authors}
\newcommand\vldbtitle{\shorttitle} 
\newcommand\vldbavailabilityurl{https://github.com/AIDASLab/SPDP}
\newcommand\vldbpagestyle{empty} 

\begin{document}
\sloppy
%\title{SPDP: Combine static and dynamic pruning with fine-grained control}
\title{Unified Static–Dynamic Pruning for Efficient LLM Inference}

%%
%% The "author" command and its associated commands are used to define the authors and their affiliations.30

\author{Jinhyeok Kim}
\affiliation{%
  \institution{Seoul National University}
  % \streetaddress{P.O. Box 1212}
  \city{Seoul}
  \state{South Korea}
}
\email{jhkim00@snu.ac.kr}

\author{Yejoon Lee}
\affiliation{%
  \institution{Seoul National University}
  % \streetaddress{P.O. Box 1212}
  \city{Seoul}
  \state{South Korea}
}
\email{leeyejoon@snu.ac.kr}

\author{Jaeyoung Do}
\authornote{Corresponding author.}
\affiliation{%
  \institution{Seoul National University}
  \city{Seoul}
  \country{South Korea}
}
\email{jaeyoung.do@snu.ac.kr}

\newcommand{\SPDP}[1]{\textsc{SPDP}}
\newcommand{\reviewbox}[1]{%
  \todo[color=orange,size=\small]{#1}%
}
\newcommand{\jae}[1]{\textcolor{black}{#1}}
\newcommand{\TODO}[1]{\textcolor{red}{#1}}
\newcommand{\yej}[1]{\textcolor{black}{#1}}
\newcommand{\jhk}[1]{\textcolor{black}{#1}}

\begin{abstract}
The increasing deployment of large language models (LLMs) has magnified the computational and memory bottlenecks of autoregressive decoding, where low compute intensity and bandwidth-bound kernels dominate inference cost. Weight pruning offers a promising remedy, but existing methods remain confined to either static pruning (SP)—which permanently removes redundant weights but lacks adaptivity—or dynamic pruning (DP)—which adapts to input sparsity but introduces runtime irregularity. This paper presents SPDP, a unified sparse-inference framework that integrates unstructured SP with input-adaptive DP for efficient LLM inference on GPUs. SPDP co-designs a new Tiled-Column-wise Bitmap Compressed (Tiled-CBC) format and two complementary GPU kernels: (1) a CUDA-core spMspV kernel featuring Hybrid Activation-aware Dynamic Shared-Memory Bitmap Decoding (HAD-SMBD) for fine-grained, runtime activation skipping, and (2) a Tensor-Core SpMM kernel optimized for prefill computation. This joint format–kernel design harmonizes static and dynamic sparsity, maintaining bandwidth-efficient memory access and high compute intensity under both phases of LLM inference. Comprehensive evaluations on inference-optimized GPUs demonstrate that SPDP achieves 1.24×–1.37× average speedup (up to 2.51×) over state-of-the-art sparse frameworks such as SpInfer, while matching perplexity with up to 25\% higher sparsity. SPDP advances the inference efficiency–quality Pareto frontier, showing that unified static–dynamic pruning can deliver substantial throughput and performance-per-watt improvements in large-scale LLM serving.
\end{abstract}

\maketitle

\pagestyle{\vldbpagestyle}
\begingroup\small\noindent\raggedright\textbf{PVLDB Reference Format:}\\
\vldbauthors. \vldbtitle. PVLDB, \vldbvolume(\vldbissue): \vldbpages, \vldbyear.\\
\href{https://doi.org/\vldbdoi}{doi:\vldbdoi}
\endgroup
\begingroup
\renewcommand\thefootnote{}\footnote{\noindent
This work is licensed under the Creative Commons BY-NC-ND 4.0 International License. Visit \url{https://creativecommons.org/licenses/by-nc-nd/4.0/} to view a copy of this license. For any use beyond those covered by this license, obtain permission by emailing \href{mailto:info@vldb.org}{info@vldb.org}. Copyright is held by the owner/author(s). Publication rights licensed to the VLDB Endowment. \\
\raggedright Proceedings of the VLDB Endowment, Vol. \vldbvolume, No. \vldbissue\ %
ISSN 2150-8097. \\
\href{https://doi.org/\vldbdoi}{doi:\vldbdoi} \\
}\addtocounter{footnote}{-1}\endgroup

\ifdefempty{\vldbavailabilityurl}{}{
\vspace{.3cm}
\begingroup\small\noindent\raggedright\textbf{PVLDB Artifact Availability:}\\
The source code, data, and/or other artifacts have been made available at \url{\vldbavailabilityurl}.
\endgroup
}

\section{Introduction} 

The rapid evolution of Large Language Models (LLMs)~\cite{touvron2023llama, openaio1, deepseekr1, yang2025qwen3technicalreport} has fundamentally transformed artificial intelligence, enabling state-of-the-art performance in reasoning \cite{openaio1, deepseekr1}, dialogue \cite{instructgpt}, summarization \cite{summ_survey}, and code generation \cite{code-survey}. However, this progress comes at substantial costs: modern LLMs with tens of billions of parameters demand immense memory bandwidth and arithmetic throughput. The problem has intensified with the advent of \emph{test-time scaling}~\cite{openaio1, deepseekr1}—which intentionally increases computation during inference to enhance reasoning quality—and the rise of \emph{agentic AI} systems \cite{agent-survey, yao2023react, schick2023toolformer} that autonomously plan, reason, and interact with tools. These paradigms drastically increase the number of generated tokens per session, shifting workloads toward iterative, decode-heavy inference where memory bandwidth and latency become critical bottlenecks, thereby reducing performance-per-watt due to the power-inefficient decoding phase~\cite{fernandez2025energyconsiderationslargelanguage}. 

To address these growing costs, model compression has re-emerged as a cornerstone for efficient LLM serving. Among various compression strategies, \emph{weight pruning} (or sparsification) has shown strong promise in reducing both memory footprint and arithmetic cost~\cite{magnitude-pruning, sparsegpt, wanda, ria, zhou2021learning}. By removing redundant weights, pruning introduces sparsity that can theoretically reduce the number of required floating-point operations. Pruning approaches differ in granularity: structured pruning eliminates entire neurons or channels, offering hardware regularity but often at the cost of accuracy; semi-structured pruning provides a middle ground that can be exploited by custom kernels; and unstructured pruning offers the finest granularity, selectively removing individual weights while typically preserving accuracy even at high sparsity levels.

\begin{figure}[t]
  \centering
  \setpipelineheight[0em]
      {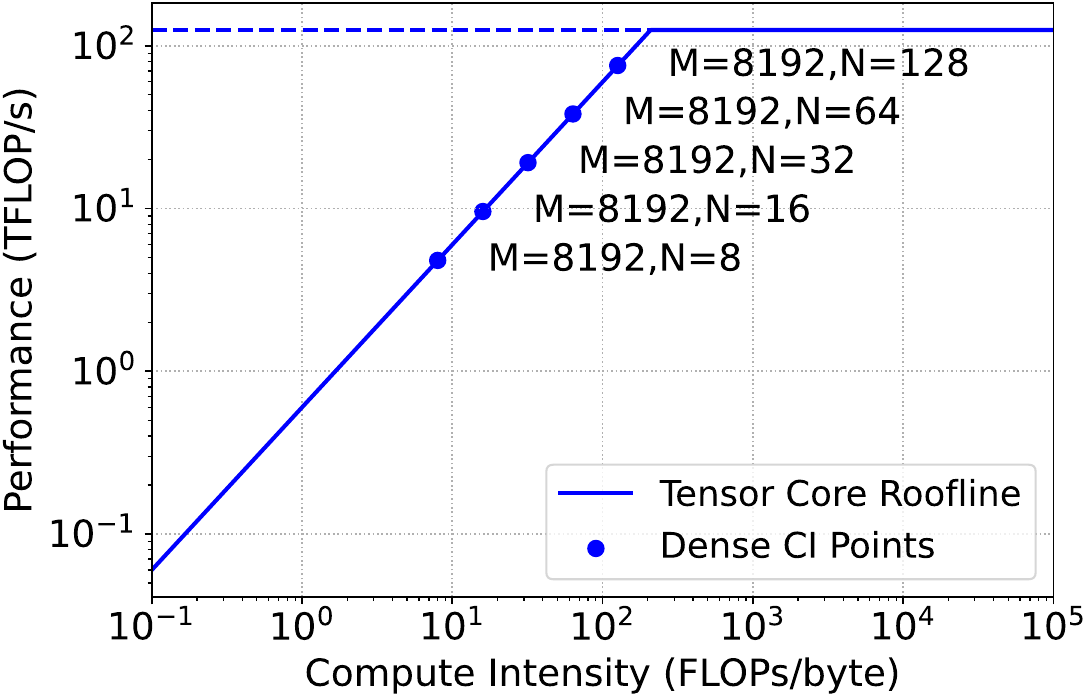}
    {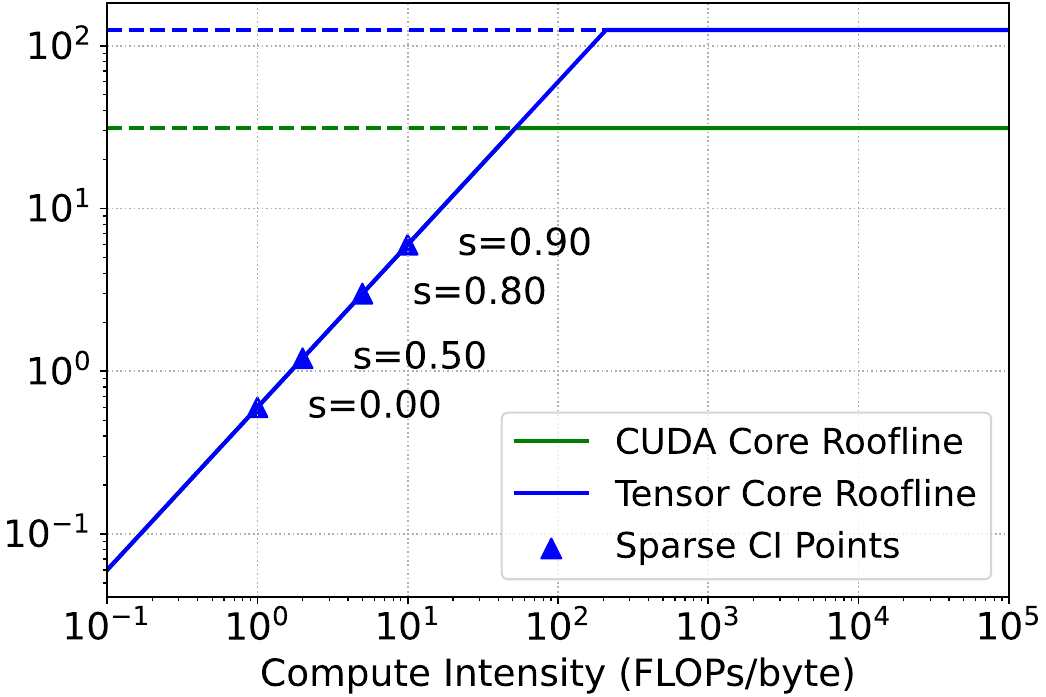}

  \makebox[\linewidth][c]{%
    \subfloat[Output with scaling in GEMM]{%
      \includegraphics[height=0.95\pipelineH]{figures_motiv/a10_roofline_dense_points_colored.pdf}%
      \label{fig:roofline_a}%
    }%
    \hspace{\pipelinegap}%
    \subfloat[Sparsity scaling in SpMV]{%
      \includegraphics[height=0.95\pipelineH]{figures_motiv/a10_roofline_sparse_points_colored.pdf}%
      \label{fig:roofline_b}%
    }%
  }

  \caption{Roofline analysis of matrix multiplication on an A10G GPU.
    (a) Effect of output width ($N$) on compute intensity and throughput in dense GEMM. As $N$ decreases, performance shifts from compute- to memory-bound. (b) Effect of sparsity ($s$) on throughput in sparse SpMV ($M=8192$, $N=1$). Higher sparsity reduces memory traffic and increases compute intensity.}
  \label{fig:roofline}
\end{figure}
    
While structured and semi-structured pruning achieve practical 
acceleration on specific hardware, their rigid sparsity patterns make it difficult to preserve model quality and generalize across architectures. As a result, recent research has increasingly focused on \emph{unstructured pruning}—the most flexible yet system-challenging paradigm—where individual weights are selectively removed without any structural constraints. Despite its algorithmic advantages, unstructured sparsity introduces substantial \emph{system-level challenges} for large-scale LLMs. Whereas smaller networks can tolerate extreme sparsity (70–90\%) with minimal accuracy degradation, large LLMs exhibit much lower redundancy and suffer pronounced quality loss at similar sparsity levels. \rb{Furthermore, autoregressive decoding reduces computation to inherently memory-bound skinny matrix multiplications, making the realized benefit of sparsity highly dependent on how the sparse structure is represented and scheduled on GPUs.} Consequently, general-purpose sparse kernels 
%such as cuSPARSE~\cite{nvidia_cusparse_2023}, as well as DNN-optimized SpMM frameworks like Sputnik~\cite{sputnik} and SparTA~\cite{sparta}, 
~\cite{nvidia_cusparse_2023, sputnik, sparta}
often fail to provide meaningful acceleration under moderate sparsity (30–60\%)—a regime typical in LLM pruning—due to limited compute intensity and nontrivial metadata overheads.

Recent systems such as Flash-LLM~\cite{flash-llm}, SpInfer~\cite{spinfer} address this limitation by introducing \emph{load-as-sparse, compute-as-dense} pipelines that reconstruct dense tiles for Tensor Core execution, effectively mitigating memory bottlenecks. However, these methods rely exclusively on \emph{static sparsity}, where pruning patterns are fixed offline, thus forfeiting adaptivity to input-dependent redundancy. In contrast, \emph{dynamic pruning}~\cite{teal, griffin, cats, dejavu, federici2025efficient} adaptively skips computation for unimportant activations at runtime, offering input-aware efficiency. Unfortunately, existing dynamic pruning frameworks typically restrict themselves to coarse-grained or structured activation masks for hardware compatibility, limiting fine-grained control and failing to exploit the potential of unstructured sparsity.

Our key insight is that static and dynamic pruning are inherently complementary. Static pruning (SP) permanently removes redundant \emph{weights}, reducing memory footprint and bandwidth cost, while dynamic pruning (DP) adaptively skips redundant \emph{activations}, reducing token-dependent computation. \rb{SP and DP act along orthogonal dimensions: SP prunes weights across rows, whereas DP skips activations across columns.} Their sparsity patterns therefore overlap minimally and \emph{multiply} effective sparsity. Our roofline-based analysis (Figure~\ref{fig:roofline}) shows that this combined sparsity significantly raises compute intensity and attainable throughput in memory-bound decode workloads, offering a new efficiency frontier for LLM inference (for more details, see Section~\ref{subsec:pruning-roofline}).

However, translating this theoretical complementarity into practical acceleration exposes \emph{system-level incompatibilities}. Existing SP and DP frameworks are not directly composable: static sparse kernels are row-major and optimized for sparse matrix–matrix multiplication (SpMM), whereas dynamic pruning induces column-wise sparsity and reduces workloads to sparse matrix–sparse vector multiplication (spMspV). This mismatch disrupts coalesced memory access, breaks warp-level stride regularity, and severely degrades GPU throughput. Furthermore, Tensor Core pipelines—designed for dense, fixed-tile fragments—cannot efficiently process per-token activation sparsity, leading to sharp drops in utilization. Therefore, achieving end-to-end acceleration under joint SP–DP sparsity requires full-stack co-design across layout, scheduling, and execution.

This paper presents \textbf{\frameworktitle{}}, a unified sparse-inference framework that integrates unstructured \emph{static pruning} with input-adaptive \emph{dynamic pruning} for efficient LLM inference. \frameworktitle{} bridges the static and dynamic paradigms through a joint \emph{format–kernel co-design}. At its core is the \emph{Tiled-Column-wise Bitmap Compressed (Tiled-CBC)} format, which encodes statically pruned weights in column-major micro-tiles aligned with DP’s activation sparsity pattern. Built atop this layout, a CUDA-core-based \emph{spMspV kernel} exploits hybrid activation-aware decoding and a fine-grained asynchronous pipeline to skip inactive columns dynamically while preserving coalesced memory access. During the prefill stage, the same format seamlessly supports a Tensor-Core-based \emph{SpMM kernel} through efficient shared-memory layout alignment, maintaining high compute utilization without reformatting. Together, these designs make it possible—\emph{for the first time}—to unify static and dynamic sparsity under a single GPU execution framework that achieves both algorithmic and system-level efficiency.

Comprehensive evaluations on inference-optimized GPUs show that \frameworktitle{} achieves higher SM utilization—the fraction of active GPU compute units—and fewer shared-memory conflicts than state-of-the-art systems such as SpInfer, while sustaining comparable memory throughput. End-to-end experiments on LLaMA-2–7B confirm that \frameworktitle{} reduces time per output token (TPOT) at matched perplexity, demonstrating that static–dynamic co-pruning improves the speed–quality trade-off in memory-bound decoding.

This paper makes the following contributions:
\begin{itemize}[leftmargin=*]
\item We present a unified analysis of static and dynamic pruning, revealing their orthogonal sparsity dimensions and quantifying their theoretical synergy through roofline modeling.
\item We propose \frameworktitle{}, a co-designed sparse inference framework that harmonizes unstructured static pruning and input-dependent dynamic pruning for efficient LLM inference.
\item We introduce the \emph{Tiled-CBC} format and two complementary sparse kernels: (1) a CUDA-core \texttt{spMspV} kernel with Hybrid Activation-aware Dynamic Shared-Memory Bitmap Decoding (HAD-SMBD), and (2) a Tensor-Core \texttt{SpMM} kernel for prefill.

\item We empirically demonstrate that combining SP and DP advances the inference efficiency–quality Pareto frontier. \frameworktitle{} achieves \textbf{1.24× and 1.37× average speedup (up to 1.70× and 2.51×)} over SpInfer, and \textbf{1.88× and 2.11× average speedup (up to 3.32× and 3.52×)} over the cuBLAS baseline on A10G and L4 GPUs, respectively. While maintaining comparable perplexity with \textbf{up to 25\% higher sparsity}, \frameworktitle{} delivers both lower latency and improved performance-per-watt in large-scale LLM serving.
\end{itemize}

\section{Background \& Related Works}
We review LLM inference and pruning-based acceleration. We contrast prefill and decode, categorize pruning by granularity and adaptivity, and summarize sparse GPU kernels for LLMs.

\subsection{LLM Inference}

Modern Transformer-based LLM \cite{transformer} inference comprises two workloads: \textit{prefill} and \textit{decode}. Prefill computes the KV cache of all prompt tokens in a single forward pass, whereas decode generates one token per pass and appends its KV states to the cache. In prefill, the input tensor has shape [B, L, H], where B, L, and H denote batch size, sequence length, and hidden dimension, respectively. Large B and L amortize the cost of loading weights into on-chip memory across both dimensions. Decode instead operates on [B, 1, H], yielding much lower compute intensity than prefill; at $B=1$, the linear layers reduce to matrix-vector multiplications.
\subsection{Pruning}

Pruning the weights of neural network can reduce the size of weights loaded to on-chip memory. Based on its granularity, pruning can be divided into two categories: (semi-) structured pruning and unstructured pruning.
(Semi-) structured pruning prunes structured elements, such as the neurons of an MLP layer \cite{ma2023, bansal-etal-2023-rethinking, dejavu, an2024} or N elements out of M consecutive elements \cite{zhou2021learning, maskllm}, while unstructured pruning \cite{sparsegpt, wanda, hubara21} prunes the individual elements. Pruning methods are also divided into two distinct flavors, based on whether pruning is dependent on the input.

\textbf{Static Pruning. }
Static pruning (SP) applies a fixed mask $\mathcal{M}$ to the weight matrix $W$, resulting in a pruned weight $W_{sp}$. $\odot$ denotes element-wise multiplication.
\[
W_{sp} = W \odot \mathcal{M}, \ W \in \mathbb{R}^{M \times K}, \mathcal{M} \in \{0,1\}^{M \times K}
\]

% \reviewtag{m:r3d3}{R3}{D3}
% \ptmark{m:r3d3}
% \MNOTE{\textbf{R3}\\D3}

Structured static pruning is the most common and easiest way to accelerate, however, it degrades the quality of LLM compared to classical deep learning models. Therefore, 2:4 semi-structured pruning emerges with NVIDIA Sparse Tensor Core (SpTC)~\cite{mishra2021accelerating}, which enhances the tradeoff between model quality and speedup. Various works aim to mitigate the quality degradation imposed by the rigid 2:4 constraint, including learnable semi-structured mask training~\cite{maskllm}, adaptive matrix reparameterization for high-performance 2:4 sparsity~\cite{liu2026armor}, proximal regularization frameworks for structured mask learning~\cite{liu2025proxsparse}, as well as hardware-aware 2:4 pruning and fine-tuning strategies designed to fully exploit SpTC. 

In this work, we limit our focus to \textit{unstructured} static pruning. Magnitude pruning \cite{magnitude-pruning} prunes individual elements based on their magnitude. SparseGPT \cite{sparsegpt} formulates the selection problem by sparse regression and proposes an efficient solve. Wanda \cite{wanda} considers activation feature norms alongside weight magnitudes. Unstructured static pruning better preserves model quality and, owing to its fine granularity, enables continuous sparsity ratios and flexible layer-wise allocation across Transformer blocks~\cite{lu2024alphapruning, chen2025dlp}. In contrast, structured pruning operates at a coarser granularity and often incurs larger accuracy degradation, while fixed-pattern semi-structured methods (e.g., 2:4 sparsity) restrict the achievable sparsity allocation despite being easier to accelerate on modern GPUs.

\textbf{Dynamic Pruning. }
Dynamic pruning (DP), also known as activation sparsity or contextual sparsity \cite{dejavu, relufication, lte, prosparse, griffin, cats, teal}, prunes the LLM based on the given input. The mask $\mathcal{M}$ is not fixed; it is instead dynamically determined by the input $x$. 
\[
W_{dp}(x) = W \odot \mathcal{M}(x), \ W \in \mathbb{R}^{M \times K}, \mathcal{M}(x) \in \{0,1\}^{M \times K}, x \in \mathbb{R}^K
\]
DejaVu \cite{dejavu} pioneered DP by exploiting the high sparsity of activations, stemming from using ReLU as an activation function. 

Recent DP methods \cite{griffin, cats, teal} extends DP to general models not using ReLU. Existing DP methods typically operate at a structured granularity, which limits opportunities for finer-grained sparsity.

To the best of authors' knowledge, DuoGPT \cite{duogpt} is the only work so far to combine dynamic pruning with static pruning. It achieves better generation quality compared to SparseGPT or Wanda, under the same compression ratio. However, the speedup from sparsity is only provided as a theoretical argument, leaving contribution for GPU kernel implementation and practical speedup.

\subsection{Unstructured Pruning Acceleration}\label{subsec:unst_accel}

Exploiting unstructured pruning for acceleration has long been constrained by the inefficiencies of sparse matrix–matrix multiplication (SpMM) on GPUs.
% Traditional GPU libraries such as cuSPARSE~\cite{nvidia_cusparse_2023} (using CSR, COO, or ELL formats) are primarily optimized for scientific workloads, where matrices are extremely sparse. Although research on GPU SpMM continues to advance~\cite{10.1145/3620666.3651378, 10.1145/3710848.3710888, nvidia_cusparse_2023, 10.5555/3703596, okanovic2024high}, these studies generally focus on such highly sparse regimes. 
Traditional GPU libraries such as cuSPARSE~\cite{nvidia_cusparse_2023}, sparse kernel generator TACO~\cite{taco}, or recent works on GPU SpMM~\cite{10.1145/3620666.3651378, 10.1145/3710848.3710888, nvidia_cusparse_2023, okanovic2024high, 11112857} or SpMV~\cite{srsparse, li2020adaptive} generally focus on extremely high sparsity, often targeting scientific workloads. In contrast, LLMs tend to exhibit moderate sparsity, where the overhead of index storage and irregular memory access often outweighs the computational savings.

Sputnik~\cite{sputnik} and SparTA~\cite{sparta} introduced DNN-oriented sparse GPU kernels. Sputnik exploited regularities of DNNs, and SparTA adopted a compiler-level design with the Tensor-with-Sparsity-Attribute (TeSA) abstraction, generating specialized kernels for known sparsity patterns. SparseTIR~\cite{ye2023sparsetir} proposed a composable sparse compilation framework that supports flexible sparse formats and schedule transformations. However, its performance remains low compute intensity in LLM workloads.

Flash-LLM~\cite{flash-llm} advanced this line of work with a hardware-aware \emph{load-as-sparse, compute-as-dense} (LSCD) approach. Instead of performing operations directly in the sparse domain, Flash-LLM loads pruned weights in compressed form from global memory, reconstructs dense tiles on-chip using shared memory, and executes them with dense Tensor Core instructions. This LSCD-based design effectively mitigates the memory-bandwidth bottleneck while exploiting the GPU’s high-throughput compute units. Building upon this, SpInfer~\cite{spinfer} further introduced a Tensor-Core-Aware Bitmap Encoding (TCA-BME) scheme, which compresses sparse matrices using lightweight bitmaps optimized for warp-level access patterns, achieving state-of-the-art SpMM performance on modern GPUs. \jhk{At the hardware-software co-design level, Coruscant~\cite{10.1145/3725843.3756065} co-designed a bitmap-based SpMM kernel with a Sparse Tensor Core extension to efficiently support unstructured LLM weight sparsity.}

\section{Motivation \& Challenges}
We examine when sparsity benefits LLM inference, focusing on memory-bound decode. A roofline analysis (Section~\ref{subsec:pruning-roofline}) shows that reducing data movement via compression and skipping raises attainable throughput. This motivates combining static and dynamic pruning, which act on orthogonal dimensions (Section~\ref{subsec:combine_sp_dp}). 
% We then outline key GPU challenges—layout mismatch, metadata overhead, and reduced Tensor Core efficiency—that drive our format–kernel co-design (Section~\ref{subsec:challenge}).
\jhk{We then outline three GPU challenges: layout mismatch, metadata overhead, and reduced Tensor Core efficiency (Section~\ref{subsec:challenge}).}

\subsection{Efficacy of Pruning: Roofline Analysis}\label{subsec:pruning-roofline}

The decode phase of small-batch LLM inference is inherently \emph{memory-bound}. Since there is limited parallelism across the batch dimension, data transfers between off-chip and on-chip memory dominate execution time. This bottleneck becomes more severe on consumer-grade GPUs—commonly used for low-batch inference—which typically rely on GDDR memory offering much lower bandwidth than HBM-equipped enterprise GPUs. The limited memory bandwidth further exacerbates the memory wall in small-batch LLM decoding. Moreover, the recent rise of test-time scaling and reasoning-oriented models~\cite{openaio1, deepseekr1} drastically increases the number of generated tokens per session, further amplifying the cost of decode computation.

%In the context of sparse LLM inference, two key metrics are commonly used to characterize the efficiency of sparse computation: the \emph{Compression Ratio (CR)} and the \emph{Compute Intensity (CI)}.  

To quantitatively analyze how sparsity alleviates the memory bottleneck, we adopt the \emph{roofline model}. Two metrics characterize the efficiency of sparse computation, including the \emph{Compression Ratio (CR)} and the \emph{Compute Intensity (CI)}.

For an FP16 weight matrix $W \in \mathbb{R}^{M \times K}$, the \emph{CR} is defined as
\begin{equation}
  CR = \frac{Stor_{dense}}{Stor_{compressed}}
      = \frac{2{B} \times M \times K}{Stor_{compressed}},
\end{equation}
where ${B}$ denotes bytes, and each FP16 element occupies $2{B}$.
Here, $Stor_{compressed}$ is the total storage size of the compressed sparse matrix, while
$Stor_{dense} = 2{B} \times M \times K$ is that of the dense FP16 weight matrix.
A higher $CR$ indicates stronger compression, reducing memory footprint and data transfer overhead.

For input matrix $X \in \mathbb{R}^{K\times N}$ and a weight matrix $W \in \mathbb{R}^{M \times K}$, the \emph{CI} is the ratio of computation to memory traffic, given by
\begin{equation}
  CI = \frac{\text{FLOPs}}{\text{Bytes transferred}} 
     = \frac{M \times N}{\frac{M}{CR} + N}
\end{equation}
A higher $CI$ implies more computation is performed per unit of memory access, improving performance in memory-bound regimes.

Based on these definitions, the \emph{roofline model} expresses the achievable kernel performance as $ P = \min(P_{\text{peak}},\; CI \times B_{\text{mem}})$,
where $P_{\text{peak}}$ denotes the peak compute throughput (FLOPs/s) and $B_{\text{mem}}$ is the available memory bandwidth (Bytes/s).  

The roofline model thus visualizes the transition between the compute-bound and memory-bound regions: kernels with low $CI$ are limited by memory bandwidth, while those with high $CI$ can approach the peak compute performance (i.e., $P_{\text{peak}}$).  

Figure~\ref{fig:roofline} illustrates this effect on an NVIDIA A10G GPU. As shown in Figure~\ref{fig:roofline}a, throughput saturates at small $N$ due to memory limitations. In Figure~\ref{fig:roofline}b, increasing sparsity effectively reduces data movement, raising $CI$ and thereby achievable throughput. In our analysis, sparsity could improve performance by up to $10\times$ under memory-bound decode workloads. Therefore, pruning emerges as an essential mechanism for accelerating LLM inference, especially in small-batch or interactive settings where memory bandwidth is the primary bottleneck.

\subsection{Bring the Best of Both Worlds: Combining Static and Dynamic Pruning}
\label{subsec:combine_sp_dp}

Static and dynamic pruning each provide unique benefits but also suffer from fundamental limitations. \emph{Static pruning (SP)}~\cite{magnitude-pruning, sparsegpt, wanda, ria, zhou2021learning} removes weights offline, permanently reducing model size and memory footprint. However, since SP is input-agnostic, it may remove parameters critical for specific contexts, leading to quality degradation. \emph{Dynamic pruning (DP)}~\cite{teal, griffin, cats, dejavu, federici2025efficient}, on the other hand, adaptively zeros out activations based on input, enabling context-aware sparsity. Yet, DP typically operates at structured granularity—pruning entire columns of the weight matrix—and thus lacks fine-grained flexibility. Moreover, since the dense weights remain loaded in memory, DP offers little benefit in terms of memory footprint and introduces runtime overhead for mask generation and control flow.

Despite their differences, SP and DP are inherently complementary.
DP's input-dependent adaptivity mitigates SP's input-agnostic nature, while SP's unstructured fine granularity compensates for DP’s coarse activation-level pruning and improves overall storage efficiency.
This complementarity naturally raises the question: \emph{Can static and dynamic pruning be jointly employed to push the Pareto frontier of speed and quality?} To answer this question, we empirically analyze how SP and DP interact using state-of-the-art pruning methods.
For SP, we adopt magnitude pruning~\cite{magnitude-pruning}, SparseGPT~\cite{sparsegpt}, and Wanda~\cite{wanda}; for DP, we use TEAL~\cite{teal}.
SparseGPT and Wanda require a small calibration set to estimate input statistics.
Following prior work~\cite{sparsegpt, wanda}, we sample 128 sequences, each with the full model context length, from the C4~\cite{c4} dataset.
To evaluate the text quality of jointly pruned models, we report perplexity on the WikiText~\cite{wikitext} dataset using Llama2-7B~\cite{touvron2023llama}.
Within this setup, we also measure the number of elements pruned by both SP and DP—referred to as the \emph{overlapping sparsity}.

\begin{figure}
  \centering
  % \subfloat[Uniform perplexity combining TEAL and Wanda ($s=0.55$)]
  \subfloat[PPL under combined pruning]
  {
    \includegraphics[width=0.47\linewidth]{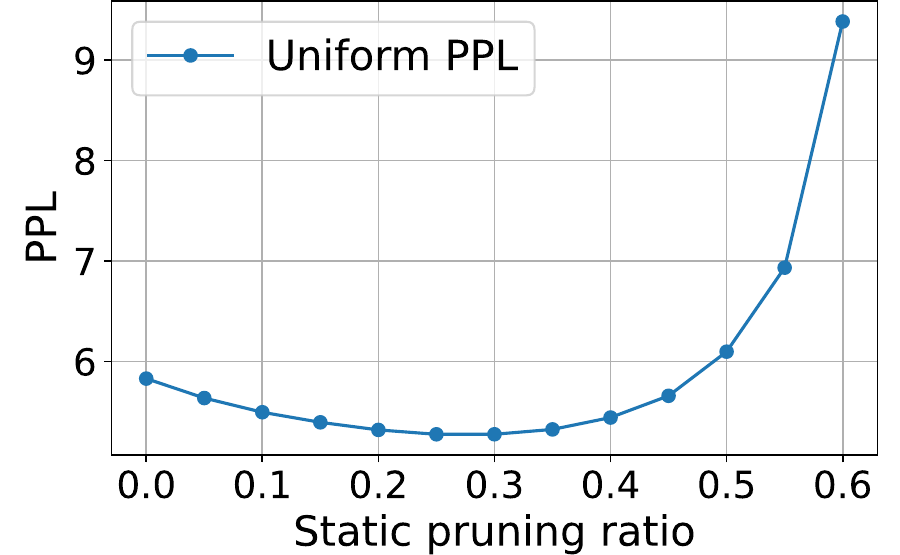}
    \label{fig:spdp_ppl0}
  }
  \hspace{-1.0em}
  \subfloat[PPL across various methods]
  {
    \includegraphics[width=0.47\linewidth]{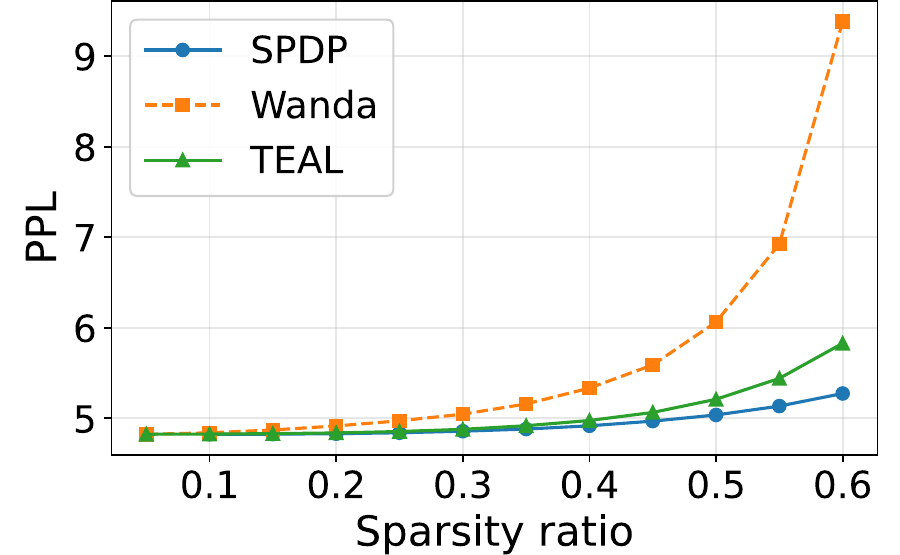}
    \label{fig:spdp_ppl1}
  }
  \caption{Perplexity comparison across pruning strategies. Combining static and dynamic pruning yields the lowest perplexity across all sparsity levels.
    }
  \label{fig:spdp_ppl}
\end{figure}

\subsubsection{Perplexity Comparison}\label{subsubsec:ppl}

We compare perplexity across various static and dynamic sparsity ratios.
When combining static sparsity $sp$ and dynamic sparsity $dp$, the total sparsity is defined as $1 -(1-sp)(1-dp)$. As shown in Figure~\ref{fig:spdp_ppl}a, the lowest perplexity is achieved when both $sp$ and $dp$ are moderate (e.g., around 0.3–0.4), rather than extreme on either side. This indicates that jointly applying SP and DP yields a balanced pruning effect—retaining essential weights while dynamically skipping unimportant activations. Figure~\ref{fig:spdp_ppl}b further shows that this advantage persists across different total sparsity levels, and becomes more pronounced at high sparsity ratios where substantial speedup is practically attainable. Overall, combining SP and DP improves the speed–quality trade-off beyond what either method can achieve alone.

\subsubsection{Overlapping Sparsity Analysis}\label{subsubsec:overlap}

When combining static and dynamic pruning, an important consideration is whether the two methods prune the same weights or act independently. If SP and DP operate independently of each other, their overlapping sparsity ratio should equal the product of the two sparsity ratios, $sp \times dp$. For example, when both $sp$ and $dp$ are 0.6, the theoretical overlap is 0.36, resulting in a total sparsity of $1 - (1-sp)(1-dp) = 0.84$. In contrast, if one pruning method fully subsumes the other, the total sparsity would remain 0.6, representing the worst-case overlap.

\begin{figure}
    \centering
    \includegraphics[width=0.95\linewidth]{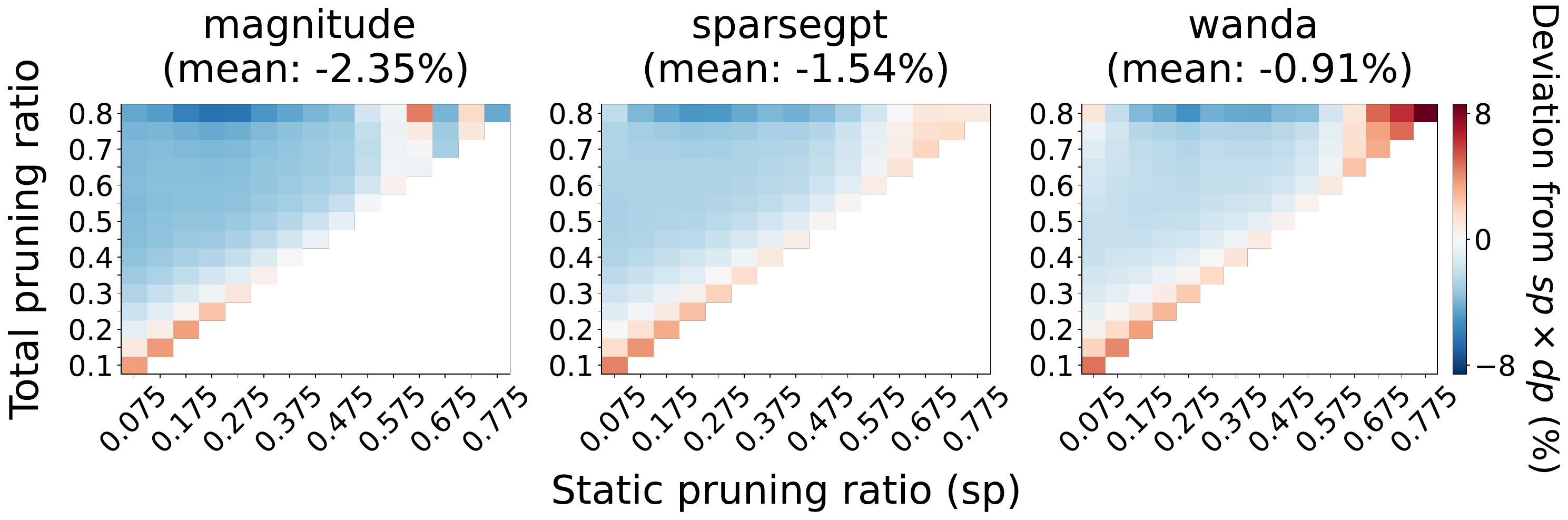}
    % \vspace{-0.5em}
    \caption{Deviation of measured overlapping sparsity from theoretical level $sp \times dp$. Color represents the deviation in \%, while X and Y axes represents the sparsity ratios. Across three SP methods combined with TEAL, overlap remains on par or slightly below the theoretical level.}
    \label{fig:spdp_overlap}
\end{figure}
% \vspace{-1em}

In the following analysis, we measure the deviation of observed overlap from the theoretical expectation. Figure~\ref{fig:spdp_overlap} shows the deviation of overlapping sparsity, as a change from the theoretical baseline $sp \times dp$. As shown in Figure~\ref{fig:spdp_overlap}, the overlaps between each unstructured SP method and TEAL remain close to or slightly below the theoretical value, indicating that SP and DP make largely independent pruning decisions. Consequently, the total sparsity $1-(1-sp)(1-dp)$ holds in practice, confirming that combining SP and DP achieves the expected level of sparsity—and thus retains the potential for proportional speedup without redundancy.
\subsubsection{Roofline Analysis}\label{subsubsec:roofline_analysis_spdp}

The theoretical benefit of combining SP and DP can be further analyzed through the roofline model introduced earlier.
We extend the formulation to capture their joint impact on CI:

\begin{equation}
CI_{SPDP} = \frac{\text{FLOPs}}{\text{Bytes transferred}}
= \frac{M \times N}{\frac{M}{CR} \cdot (1 - s) + N}
\end{equation}
where $s$ denotes the \emph{effective dynamic sparsity ratio}—the fraction of weight matrix columns skipped at runtime due to DP.
This reflects all columns removed dynamically, including those already pruned statically (i.e., overlapping sparsity).
% This ratio reflects all columns removed dynamically, including those already pruned statically (i.e., overlapping sparsity).
Accordingly, $(1 - s)$ represents the fraction of activations that remain active and thus determine the amount of weight data that must be accessed during inference.

In this formulation, SP increases the compression ratio ($CR$), while DP further reduces the number of active columns by $(1-s)$.
Together, these effects multiplicatively boost $CI$, leading to higher throughput under memory-bound conditions.
In practice, minor overlap between SP and DP may slightly reduce the effective gain, since some columns pruned by DP may already have been removed by SP.
However, as shown in Section~\ref{subsubsec:overlap}, the measured overlap remains negligible, validating that $s$ provides a reliable estimate of the runtime sparsity.

This roofline analysis provides a simple yet intuitive perspective on how SP (via $CR$) and DP (via $s$) complement each other to enhance computational efficiency, particularly in memory-bound scenarios. While the roofline model assumes ideal integration between static and dynamic sparsity, achieving such synergy in practice remains challenging due to memory layout mismatches and runtime scheduling overheads—issues further discussed in the next section.

\subsection{Challenges in Realizing Acceleration}
\label{subsec:challenge}
\noindent\rb{Although static pruning (SP) and dynamic pruning (DP) are complementary in theory, their theoretical sparsity benefit does not directly translate into practical GPU acceleration. Realizing this benefit requires system support across three coupled aspects: a sparse layout that preserves static compression under input-dependent dynamic sparsity, a decode execution path that avoids Tensor-Core underutilization for DP-induced spMspV, and prefill support that preserves Tensor-Core SpMM efficiency without maintaining a separate sparse representation. Existing unstructured SP formats~\cite{flash-llm, spinfer} and DP methods~\cite{teal} do not jointly satisfy these requirements. The resulting system-level challenges are summarized as follows.}

\textbf{(1) Layout mismatch and runtime sparse-fragment access overhead.}
\rb{SP kernel~\cite{flash-llm, spinfer} compresses a static sparse weights and optimize data movement for SpMM. DP changes the problem during decode: zeroing token-dependent activation turns it into input-dependent spMspV, where only weight columns associated with nonzero activation entries matter for the current token.}

% \rb{SP kernel~\cite{flash-llm, spinfer} compress a static sparse weight matrix and optimize data movement for SpMM. DP changes the problem during decode: by zeroing out token-dependent activation entries, it turns the operation into input-dependent spMspV, where only the weight columns associated with nonzero activation entries are useful for the current token.}

\rb{This creates a layout mismatch between SP and DP. A fixed SpMM-oriented format may still decode weight fragments that are inactive under DP. Avoiding this waste requires locating only the active sparse fragments at runtime, but unstructured SP makes those fragments irregular because each column contain different number of nonzeros. Therefore, realizing the theoretical benefit of combined SP and DP requires runtime-aware sparse access that preserves static compression while respecting input-dependent activation sparsity.}

\textbf{(2) Hardware-level conflict between Tensor-Core utilization and DP. }
\rb{Tensor Cores excel at matrix--matrix computation (GEMM, SpMM), but DP reduces each token’s work to \emph{GEMV-like} (or spMspV) execution with no reuse along the $N$ dimension. Under PTX instruction \texttt{mma.m16n8k16}, CI can drop by up to $8\times$, sharply lowering Tensor-Core efficiency.
On inference-oriented GPUs~\cite{a10g, l4, l40s}, achievable Tensor-Core throughput for GEMV is only about one-fourth of the CUDA core’s FP32 peak, so utilization can collapse and throughput may fall below dense cuBLAS baselines. Falling back to CUDA cores restores fine-grained, index-aware execution, but requires a redesigned pipeline to maintain instruction-level parallelism (ILP) and overlap metadata, memory, and compute for DP-driven spMspV.}

\textbf{(3) Prefill compatibility under the same format. }
\rb{DP is applied only during decode in \frameworktitle{}; the prefill phase therefore relies on SP alone and corresponds to SpMM, which is amenable to Tensor-Core execution~\cite{flash-llm, spinfer}. A format optimized for decode-time spMspV is not automatically compatible with Tensor-Core SpMM: decode favors selective sparse-fragment access induced by runtime activation sparsity, whereas Tensor-Core SpMM requires regular MMA-friendly operand tiles. A separate prefill format would increase storage and preprocessing overhead, while a single format without careful kernel support can underutilize Tensor Cores.}

Together, these issues highlight the gap between theoretical sparsity gains and practical acceleration. Bridging it requires a framework that (i) unifies SP and DP into a bandwidth-efficient layout, (ii) enables efficient CUDA-core spMspV for decode, and (iii) preserves Tensor-Core efficiency for prefill. The next section presents the design of \frameworktitle{}, which achieves these goals via joint format--kernel co-design.

\begin{figure*}[t]
  \centering
  \setpipelineheight[2.5em]
    {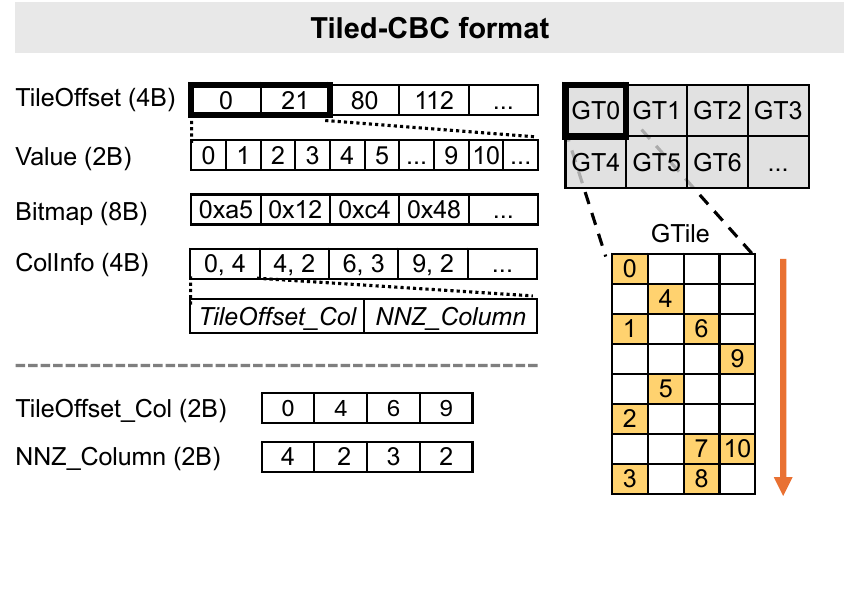}
    {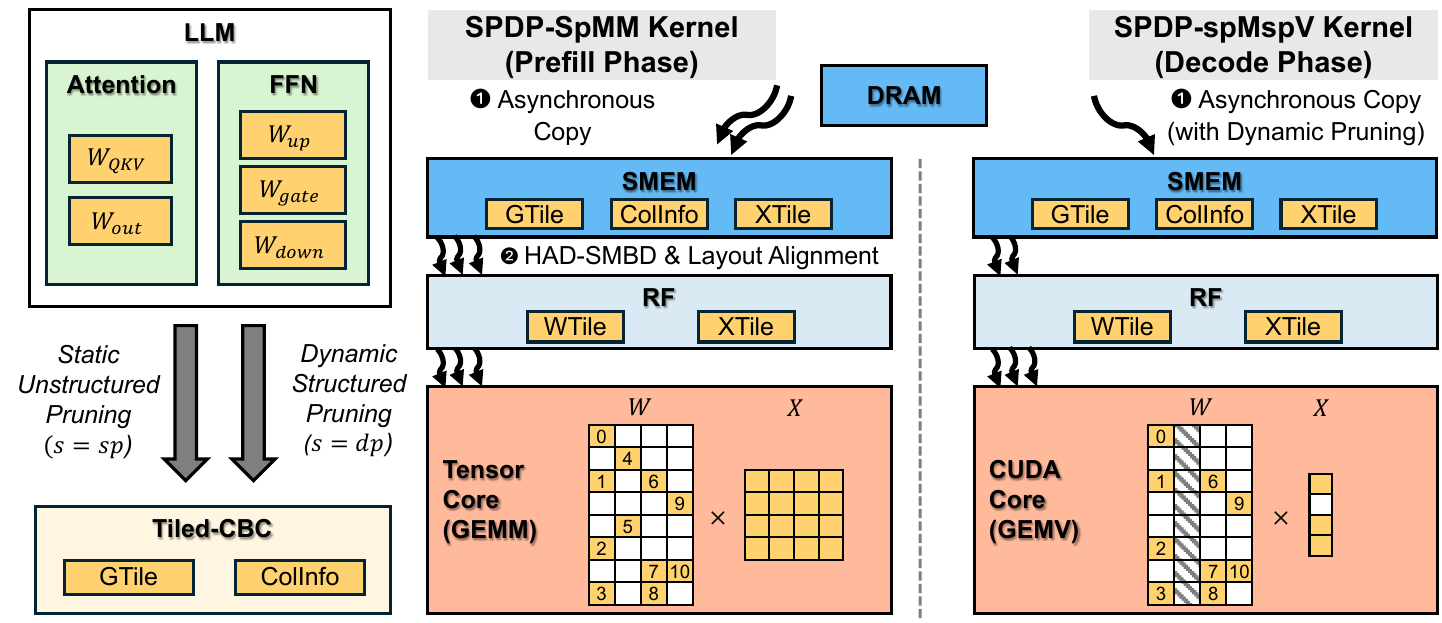}

  \makebox[\linewidth][c]{%
    \subfloat[Overview of the Tiled-CBC format.]{%
      \includegraphics[height=0.95\pipelineH]{figures_design/compression_format.pdf}%
      \label{fig:tiling_design}%
    }%
    \hspace{\pipelinegap}%
    \subfloat[Design overview of \frameworktitle{}. ]{%
      \includegraphics[height=0.95\pipelineH]{figures_design/design_overview.pdf}%
      \label{fig:design_overview}%
    }%
  }

  \caption{(a) Tiled-CBC stores nonzeros within each GTile in column-major order using a hierarchical layout. Each GTile has size $TILE\_M \times TILE\_K$, illustrated as $8 \times 4$ for simplicity. (b) \frameworktitle{} combines static pruning (SP) and dynamic pruning (DP) with the Tiled-CBC format and phase-specific inference kernels. The compressed weights are processed by the SpMM kernel during prefill and by the spMspV kernel during decode.}
  \label{fig:design_placeholder}
\end{figure*}

\section{Design of SPDP: Unifying Static and Dynamic Sparsity}
\label{sec:design_spdp}

\rb{The challenges above motivate a joint format-kernel co-design rather than a standalone sparse format or a single sparse kernel. \frameworktitle{} addresses them by combining the \textit{Tiled-CBC format} with phase-specific inference kernels, as illustrated in Figure~\ref{fig:design_overview}. Tiled-CBC (Section~\ref{subsec:tiled_cbc}) organizes statically compressed weights for efficient decode-time spMspV while retaining low indexing overhead. The CUDA-core \textit{spMspV kernel} (Section~\ref{subsec:spmspv_kernel}) then exploits this format to process input-dependent sparse activations during decode using optimized data movement, Hybrid Activation-aware Dynamic Shared Memory Bitmap Decoding (HAD-SMBD), and a fine-grained asynchronous pipeline. The Tensor-Core \textit{SpMM kernel} (Section~\ref{subsec:prefill-spmm}) supports prefill by reusing the same compressed format and arranging decoded tiles into Tensor-Core-friendly operand layouts. }
% Together, these components support efficient inference under both static and dynamic sparsity.

% \vspace{-1.5em}
\subsection{Tiled-CBC: Unified Compression Format}
\label{subsec:tiled_cbc}
% We propose a new \emph{Tiled-Column-wise Bitmap Compressed (Tiled-CBC)} format as a core component of \frameworktitle{}, designed to efficiently store statically pruned sparse matrices while seamlessly supporting DP. This Tiled-CBC format minimizes indexing overhead, enabling fine-grained activation skipping on GPUs with high compression efficiency and computational throughput. 
\rb{Motivated by the layout mismatch between static compression and dynamic access, we propose the \emph{Tiled-Column-wise Bitmap Compressed (Tiled-CBC)} format as the core sparse representation of \frameworktitle{}. Tiled-CBC makes sparse weight fragments directly addressable at column granularity while preserving a tiled representation for GPU execution, enabling decode-time spMspV to skip activation-inactive fragments with low indexing overhead.}

\textbf{Tiling Design.}
Tiled-CBC adopts a tile-level compression scheme organized in a column-major layout, since modern DP methods apply column-wise pruning along the input dimension; this layout therefore better preserves GPU-friendly memory access under DP. As illustrated in Figure~\ref{fig:tiling_design}, the weight matrix is hierarchically divided into \emph{GTiles}, each consisting of multiple column tiles.
Each column tile, with dimensions $TILE\_M\times1$ (set to $256\times1$), serves as the smallest unit for compression and pruning.
This granularity is intentionally aligned with CUDA’s 64-bit word size so that sparsity in each column can be encoded using four \texttt{uint64\_t} bitmaps, where each bit indicates whether the corresponding element is nonzero. 
Such organization ensures efficient column-wise pruning and contiguous memory access during DP, consistent with the principle of activation sparsity emphasized in TEAL~\cite{teal}.
At the outer level, each \emph{GTile} covers $TILE\_M \times TILE\_K$ elements and contains multiple column tiles. Thread blocks are assigned to process GTiles, while warps within each block handle column tiles in parallel. GTiles are stored in row-major order to preserve inter-tile coalescing and balance GPU workload distribution.

\textbf{Storage Structure.}
Tiled-CBC represents a sparse matrix with five arrays: \texttt{TileOffset}, \texttt{Values}, \texttt{Bitmap}, \texttt{ColInfo}, and optional alignment metadata. 
\texttt{TileOffset} indexes each GTile, and \texttt{Values} stores nonzeros in tile-column order.
The \texttt{Bitmap} array holds four 64-bit integers per column in the tile, indicating the location of non-zero entries.  
To facilitate DP, \frameworktitle{} introduces a compact \texttt{ColInfo} array that packs two 16-bit fields: the upper bits store the column’s starting offset within the tile (\texttt{TileOffset\_Col}), and the lower bits store the number of nonzeros in that column (\texttt{NNZ\_Column}).  
This packing enables low-overhead skipping of inactive columns at runtime, achieving contiguous memory access even under input-dependent pruning.

Let $NGT = (M/TILE\_M) \times (K/TILE\_K)$ denote the number of GTiles, $NBM = (M\times K/64)$ the number of bitmaps, and $NNZ = M \times K \times (1-s)$ the number of non-zero elements with static sparsity $s$.  
Then the memory cost of Tiled-CBC is:
\begin{align*}
Stor_{\text{Tiled-CBC}} =& 4B\times(NGT+1) + 4B\times(NGT\times TILE\_K) \\
&+ 8B\times NBM + 2B\times NNZ.
\end{align*}
% Here, each \texttt{4B} integer in \texttt{TileOffset} and \texttt{ColInfo}, \texttt{8B} bitmap per column units, and FP16 \texttt{2B} value contribute to the total storage.
Here, a 4B integer in \texttt{TileOffset} and \texttt{ColInfo}, an 8B \texttt{Bitmap} per \texttt{uint64\_t}, and an 2B \texttt{Values} each contribute to the total storage.

\rb{Compared with prior formats optimized primarily for statically pruned weights~\cite{flash-llm, spinfer}, Tiled-CBC exposes the sparse fragments needed by input-dependent DP without scanning inactive fragments. This structure preserves compression at moderate sparsity levels and provides the column-addressable layout used by the CUDA-core \frameworktitle{}-\texttt{spMspV} kernel.}
\begin{algorithm}[!t]\scriptsize

\caption{\frameworktitle{}-\texttt{spMspV} kernel pseudocode.}
\label{alg:spmspv_algorithm}
\begin{algorithmic}[1]
\State \textbf{Inputs:} \parbox[t]{.82\linewidth}{%
SparseMatrix $W$, Column Metadata $ColInfo$ (Tiled-CBC), Vector $X$, Split\_K, DP threshold}
\State \textbf{Output:} \parbox[t]{.82\linewidth}{%
Vector/Matrix $Y$ in ReductionWorkspace}

\State \textcolor{blue}{int} $BatchID = blockIdx.x;$
\State \textcolor{blue}{int} $TileY = blockIdx.y \% (M / TILE\_M),\ TileX = 0;$
\State \textcolor{blue}{int} $NumIter = ComputeIterations(BatchID, Split\_K);$

\State \_\_shared\_\_ $ValueBuffer[TILE\_M][TILE\_K];$ \Comment{\textcolor{teal}{Sparse tile buffer}}
\State \_\_shared\_\_ $BitmapBuffer[TILE\_M][TILE\_K];$ \Comment{\textcolor{teal}{Bitmap buffer}}
\State \_\_shared\_\_ $ColInfoBuffer[2][TILE\_K];$ \Comment{\textcolor{teal}{Double buffer (metadata)}}
\State \_\_shared\_\_ $XTileBuffer[2][TILE\_K];$ \Comment{\textcolor{teal}{Double buffer (dense)}}

\State \textcolor{brown}{// Prologue: Pre-loop initialization.}
\State $ColInfoLoading(ColInfoBuffer[0], ColInfo);$
\State $XTileLoading(XTileBuffer[0], X + BatchID \times TILE\_K);$ \Comment{\textcolor{teal}{Load data for DP}}
\State $cp\_async\_group\_commit();$
\State $cp\_async\_wait\_group(0);$ \Comment{\textcolor{teal}{Wait for ColInfo, XTile}}
\State $GTileLoading(BitmapBuffer, ValueBuffer, W);$ \Comment{\textcolor{teal}{Load one GTile}}
\State $cp\_async\_group\_commit();$ 
\State $ColInfoLoading(ColInfoBuffer[1], ColInfo+TILE\_K);$
\State $XTileLoading(XTileBuffer[1], X + (BatchID+1) \times TILE\_K);$
\State $cp\_async\_group\_commit();$
\State $cp\_async\_wait\_group(1);$ \Comment{\textcolor{teal}{Wait for GTile}}
\State $W_{frag} = HAD\_SMBD(ValueBuffer, BitmapBuffer);$ \Comment{\textcolor{teal}{Asynchronous decode}}
\State \_\_syncthreads();
\State \textcolor{brown}{// Main computation loop. Stages are pipelined as shown in Figure~\ref{fig:pipeline}.}
\For{\textcolor{blue}{int} $k = 0;$ $k < NumIter - 1;$ $k++$}
    \State $GTileLoading(BitmapBuffer, ValueBuffer, W + offset(k));$
    \State $cp\_async\_group\_commit();$
    \State $ColInfoLoading(ColInfoBuffer, ColInfo + offset(k));$
    \State $cp\_async\_group\_commit();$

    \State $Y_{frag} = GEMVCompute(W_{frag}, X_{frag}, Y_{frag});$\Comment{\textcolor{teal}{GEMV on CUDA cores}}

    \State $cp\_async\_wait\_group(1);$ \Comment{\textcolor{teal}{Wait for XTile, GTile}}
    \State \_\_syncthreads();
    \State $XTileLoading(XTileBuffer, X + offset(k));$
    \State $cp\_async\_group\_commit();$
    \State $W_{frag} = HAD\_SMBD(ValueBuffer, BitmapBuffer);$
    \State $cp\_async\_wait\_group(0);$ \Comment{\textcolor{teal}{Wait for ColInfo, XTile}}
    \State \_\_syncthreads();
\EndFor

\State \textcolor{brown}{// Epilogue: final iteration.}
\State $Y_{frag} = GEMVCompute(W_{frag}, X_{frag}, Y_{frag});$
\State $StoreResults(ReductionWorkspace, Y_{frag});$
\end{algorithmic}
\end{algorithm}

% \vspace{-0.1em}
\subsection{Decode Phase (spMspV) Kernel Design}
\label{subsec:spmspv_kernel}

% \rb{The hardware-level conflict between Tensor-Core utilization and DP motivates a CUDA-core decode path rather than underfilled Tensor-Core execution. Building on the column-addressable sparse fragments exposed by Tiled-CBC, we develop the \frameworktitle{}-\texttt{spMspV} kernel for decode-time execution. The kernel prefetches column metadata, loads only activation-relevant sparse fragments, decodes bitmap-compressed columns with \emph{Hybrid Activation-aware Dynamic Shared-Memory Bitmap Decoding (HAD-SMBD)}, which is further developed from SMBD of SpInfer~\cite{spinfer} and accumulates the decoded fragments with CUDA-core GEMV.}
\rb{The hardware-level conflict between Tensor-Core utilization and DP motivates a CUDA-core decode path rather than Tensor-Core execution. Building on the column-addressable sparse fragments exposed by Tiled-CBC, we develop the \frameworktitle{}-\texttt{spMspV} kernel for decode-time execution. The kernel prefetches column metadata, loads only activation-relevant sparse fragments, and decodes bitmap-compressed columns with \emph{Hybrid Activation-aware Dynamic Shared-Memory Bitmap Decoding (HAD-SMBD)}, a DP-aware extension of SMBD mechanism~\cite{spinfer}. The decoded fragments are then accumulated with the corresponding activation values using CUDA-core GEMV.}

During each iteration, a thread block performs five main procedures:
(1) \emph{ColInfo loading.} To support DP, the kernel uses \texttt{ColInfo} to determine where to prune. All threads in the block cooperatively load a \texttt{ColInfo} segment from global memory into shared memory. This metadata contains per-column offsets and nonzero counts.
(2) \emph{XTile loading.} Many DP methods require input activations for threshold comparison~\cite{teal,federici2025efficient,cats}; therefore, the kernel loads \texttt{XTile} before the corresponding weight tile. The dense input-vector tile (\texttt{XTile}) of $X^T$ is loaded into shared memory and reused without pruning, since pruning such a small vector provides negligible bandwidth savings.
(3) \emph{GTile loading.} The kernel then loads a \texttt{GTile} from global memory into a shared-memory buffer (\texttt{WTile}). Using \texttt{ColInfo} and activation information from \texttt{XTile} with a pruning metric (e.g., magnitude), the kernel selectively fetches only relevant columns from global memory to realize DP.
(4) \emph{WTile decoding.} The \texttt{WTile} in shared memory is decoded into registers using HAD-SMBD, which interprets the bitmap layout with activation-aware column skipping.
(5) \emph{CUDA core computation.} Finally, CUDA cores perform matrix--vector multiplication between the decoded \texttt{WTile} and the \texttt{XTile}, both kept in registers. This load-as-sparse, compute-as-dense (LSCD) design is motivated by prior work~\cite{flash-llm,spinfer}.

\begin{figure*}[!t]
  \centering
  \subfloat[\frameworktitle{-\texttt{spMspV} kernel Asynchronous Pipeline Design}]
    {
        \includegraphics[height=2.55cm]{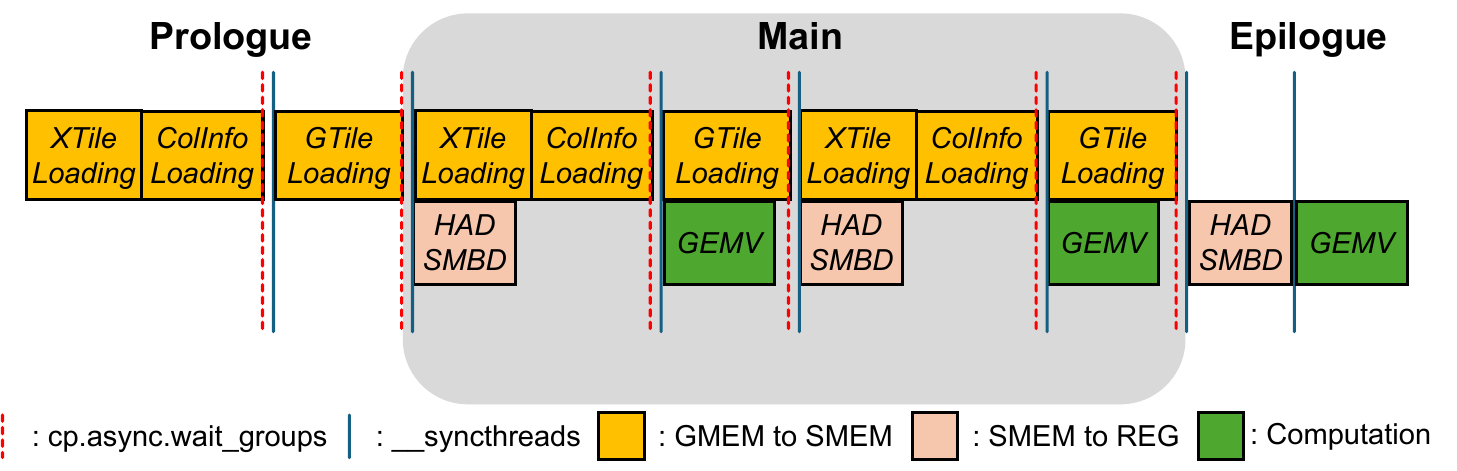}
        \label{fig:spmspv_pipeline}
    }
    \hspace{0.1em}
  \subfloat[\frameworktitle{-\texttt{SpMM} kernel Asynchronous Pipeline Design}]
    {
        \includegraphics[height=2.55cm]{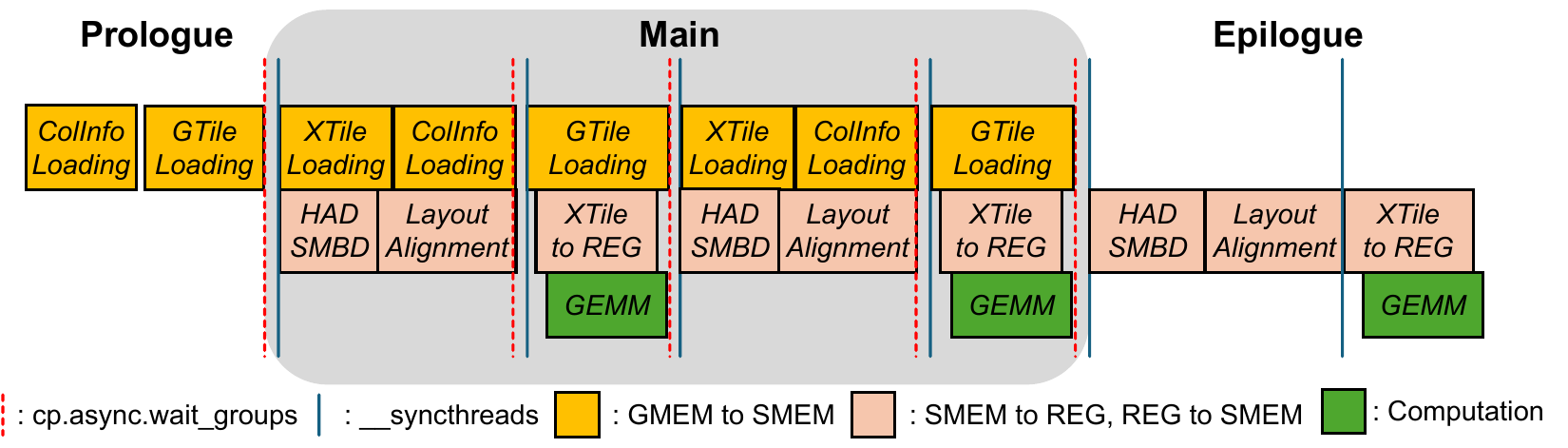}
        \label{fig:spmm_pipeline}
    }
  \caption{Asynchronous pipeline design for \frameworktitle{}-\texttt{spMspV} and \frameworktitle{}-\texttt{SpMM} kernels.
    Both kernels use HAD-SMBD for efficient decompression and fine-grained asynchronous group management to fully overlap stages without stalls. The \frameworktitle{}-\texttt{SpMM} kernel adds a Layout Alignment stage for Tensor Core execution, carefully scheduled to minimize overhead.}
  \label{fig:pipeline}
\end{figure*}

% Summarized
\subsubsection{Efficient Decompression Process}
Flash-LLM and SpInfer introduced an extract stage and Shared Memory Bitmap Decoding (SMBD) for efficient decoding under static sparsity, but they do not address decoding with input-dependent dynamic sparsity. To support the coexistence of static and dynamic sparsity, we extend SMBD from SpInfer into \emph{Hybrid Activation-aware Dynamic Shared Memory Bitmap Decoding (HAD-SMBD)}, which decodes bitmap-compressed tiles while applying dynamic activation masks.

\textbf{Register layout.}
HAD-SMBD materializes each compressed tile into warp-level register fragments using a \emph{column-major} mapping. This column-major layout matches CUDA-core-based GEMV accumulation and aligns the decoded operands with the column-oriented Tiled-CBC format.

\textbf{Hybrid two-phase decoding.}
For each column, HAD-SMBD first checks whether the column is active under dynamic pruning; inactive columns are skipped entirely.
For an active column, threads interpret its 64-bit bitmap and unpack packed nonzeros from \texttt{Value} using the per-column base offset \texttt{TileOffset\_Col} (from \texttt{ColInfo}).
In \textbf{Phase I}, each thread computes a per-thread prefix rank (the number of preceding 1-bits before its position) via \emph{MaskedPopcount}, and if its bit is set, loads the corresponding element from \texttt{Value[base + rank]} into the first FP16 lane of its register (otherwise, it inserts zero).
In \textbf{Phase II}, the thread decodes the paired FP16 lane within the same register by reusing \texttt{base} and the Phase-I rank, updating the index with only a lightweight adjustment instead of recomputing a full \emph{MaskedPopcount}.
\rb{Both phases operate on the same column, so \texttt{ColInfo} is fetched once and reused across the two phases.}

\textbf{MaskedPopcount.}
To compute the per-thread rank within the packed \texttt{Value} buffer, HAD-SMBD uses \emph{MaskedPopcount}, i.e., a prefix count of set bits in the bitmap (implemented with \texttt{\_\_popcll()}).

\textbf{Differences from SMBD.}
Compared to SMBD, HAD-SMBD (1) adopts column-major decoding for CUDA-core GEMV (instead of Tensor-Core-oriented row-major fragments), (2) leverages \texttt{ColInfo} to obtain column offsets required by the Tiled-CBC layout, and (3) keeps this shared-memory metadata access efficient (and skippable for DP-inactive columns), enabling DP-compatible decoding without disrupting the static bitmap pipeline.

Overall, HAD-SMBD bridges static bitmap decoding (for SP) with activation-dependent column skipping (for DP) in a unified decompression path.

\subsubsection{Asynchronous Pipeline Design}

We further optimized a fine-grained asynchronous pipeline from SpInfer, to eliminate the overhead while benefiting from memory access enabled by DP. Figure~\ref{fig:spmspv_pipeline} shows the asynchronous pipeline design, which improves the utilization of compute by maximizing the overlap between memory transfers and computation operation.

\textbf{Fine-Grained Asynchronous Group Management.}
\frameworktitle{} adopts double buffering with two shared-memory buffers for GTiles and XTiles, allowing the next tiles to be prefetched while the current ones are being processed.
Unlike SpInfer, which uses two asynchronous copy groups, we introduce a third independent group for \textit{ColInfo} loading, enabling three concurrent operations: (1) metadata prefetching for DP decisions (\emph{ColInfo loading}), (2) \emph{XTile loading} for upcoming computations, and (3) \emph{WTile decoding} via Hybrid Activation-aware Shared Memory Bitmap Decoding (HAD-SMBD).
Since these tasks are independent, their concurrent execution fully hides the latency of HAD-SMBD and maintains high instruction-level parallelism with minimal synchronization.

\textbf{Prefetching by Metadata Prefetching.}
Beyond overlapping weight and activation loading, \frameworktitle{} adds a preemptive \emph{metadata prefetching} stage that retrieves column activation patterns ahead of data loading.
The prefetched metadata identifies which columns of GTile should be fetched in the next iteration, forming a two-step prefetch pipeline—metadata first, data second—so that only activation-relevant tiles are transferred. 
This metadata-driven prefetching ensures continuous overlap between decoding and memory access phases, eliminating unnecessary memory transactions and maximizing memory–compute concurrency. 
Although the added prefetching and double buffering slightly increase shared memory usage and extend the pipeline’s prologue and epilogue, their effect is negligible. This is because occupancy is already bounded by register usage, and the main loop typically runs for many iterations, making the relative cost of the prologue and epilogue latency insignificant. 
Consequently, \frameworktitle{} executes dynamic pruning with near-zero additional latency while preserving the lightweight, static-pipeline structure of SpInfer.

\subsubsection{Implementation details}
% We implemented our kernel to effectively utilize the hardware, and several implementation details are described below.
\jhk{We further optimize our kernel to effectively utilize the underlying hardware, as detailed below.}

Throughout steps (1)–(3), \frameworktitle{} employs \texttt{cp.async} (compiled to \texttt{LDGSTS.128}) for high-efficiency global-to-shared memory transfer.
Introduced in the Ampere architecture~\cite{a100_whitepaper}, this mechanism streams data directly from global to shared memory, reducing latency and register pressure.
These transfers can be committed and synchronized using their intrinsic instructions, as illustrated in Algorithm~\ref{alg:spmspv_algorithm}.
Each thread transfers a 128-bit vector (eight FP16s), and arrays are padded to satisfy alignment requirements.
In particular, each column’s values are padded to 16-byte boundaries (e.g., if 7 FP16 values exist, 2 bytes are padded), which is negligible with the large $TILE\_M=256$.
Given the large $TILE\_M$, we set $TILE\_K=16$ to balance per-iteration granularity and overhead.

In step (3), the kernel streams a \texttt{GTile} into shared memory with an activation-aware selection pipeline.
Using magnitude-based DP (e.g., TEAL), threads determine active columns and access \texttt{ColInfo} metadata from shared memory to compute offsets and nonzero counts without extra global reads.
Predicate registers guard inactive columns to avoid warp divergence, while active ones proceed with coalesced \texttt{LDGSTS.128} transactions.
The column loop is unrolled to enhance ILP and reduce loop overhead.
We further adopted a tile-based scheme combined with Split-K parallelization to evenly distribute workloads, where each thread block processes an independent segment of the $K$ dimension.

\subsection{Prefill Phase (SpMM) Kernel Design}\label{subsec:prefill-spmm}

While \frameworktitle{} is highly efficient for spMspV in decode, prefill must also be supported to cover the entire pipeline of LLM inference. Prefill, where only SP is applied, involves \emph{sparse matrix-matrix multiplication (SpMM)}. However, this conflicts with our Tiled-CBC format, primarily designed for spMspV.

Therefore, we designed the \frameworktitle{}-\texttt{SpMM} kernel for prefill, as shown in Figure~\ref{fig:spmm_pipeline}, while maintaining minimal overhead from Tiled-CBC format. The overall design of the \frameworktitle{}-\texttt{SpMM} kernel is similar to that of \frameworktitle{}-\texttt{spMspV} kernel, but differs in several stages.

\rb{First, prefill corresponds to matrix--matrix multiplication and is therefore amenable to Tensor-Core execution; we leverage \emph{Tensor Cores}, following prior work~\cite{spinfer,flash-llm}. }
Second, we introduce an additional \textit{layout alignment stage} to align the register layout between the Tiled-CBC format and the Tensor Core format. Third, we eliminate the DP stage to reduce overhead of it, mainly from \emph{WTile, GTile loading}. Finally, several stages, including \textit{XTile loading} and \textit{XTile-to-register transfer}, are modified or newly added to accommodate the transition from vector to matrix input. 

\subsubsection{Layout Alignment.}
The Layout Alignment stage converts the sparse weight tile (\texttt{WTile}) stored in the \frameworktitle{}-\texttt{spMspV} layout (column-major) into the Tensor-Core-compatible operand layout (row-major) required for SpMM.
The conversion is performed entirely in shared memory, which serves as a low-latency staging buffer.
However, a naive transpose-like reordering can cause severe shared-memory bank conflicts because threads within a warp access non-contiguous addresses.

To avoid this penalty, we apply a padding-based scheme.
Each thread stores its elements into shared memory with a \texttt{TILE\_K\_PAD} stride (i.e., with an extra padded column per bank segment), which offsets addresses across lanes so that the threads in a warp map to distinct banks.
The data is then reloaded into registers following the Tensor Core operand mapping, resulting in a bank-conflict-free transpose and improved alignment with \texttt{ldmatrix.x4} loads.
% In the second pass, data is reloaded into registers following the Tensor Core operand mapping, resulting in a bank-conflict-free transpose and improving alignment with \texttt{ldmatrix.x4} loads.

This two-phase reordering introduces only a single \texttt{\_\_syncthreads()} barrier and some additional shared-memory traffic, but the overhead is negligible compared to the latency avoided by eliminating bank conflicts.
The padding cost is also minimal because occupancy is constrained by register usage, and the stage reuses the shared-memory buffer from the previous stage without requiring additional shared-memory allocation.

\subsubsection{Asynchronous pipeline design}
Similar to the \frameworktitle{}-\texttt{spMspV} kernel, the \frameworktitle{}-\texttt{SpMM} kernel uses an asynchronous pipeline, but it additionally accounts for the layout-alignment stage. Compared with decode-time \texttt{spMspV}, prefill removes DP-specific column selection in GTile loading stage and reduces the overhead of HAD-SMBD, while introducing layout alignment for Tensor-Core SpMM. \rb{In prefill, XTile spans hundreds to thousands of prompt tokens, increasing the latency of the XTile loading stage compared with decode-time \frameworktitle{}-\texttt{spMspV}. This longer stage provides a larger window to overlap activation loading with sparse-tile decoding, layout alignment, and Tensor-Core computation.}

Although \frameworktitle{}-\texttt{SpMM} primarily targets prefill, it can be extended to support batched DP in both prefill and decode; doing so would require kernel modifications and additional algorithmic support.
\begin{figure*}[t]
% \begin{figure*}[!tH]
  \centering
  \subfloat[Accuracy on downstream tasks]{
    \includegraphics[height=2.4cm]{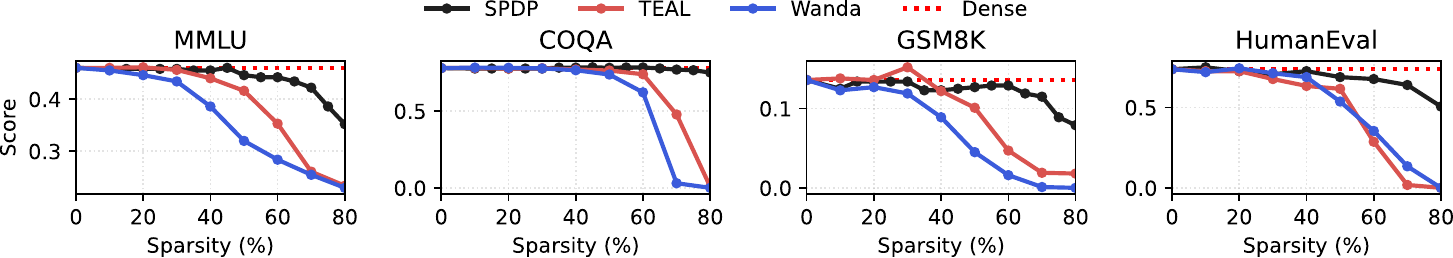}
    \label{fig:downstream_task_result_a}
  }
  \hspace{-0.7em}
  \subfloat[Compatibility with quantization]{
    \includegraphics[height=2.4cm]{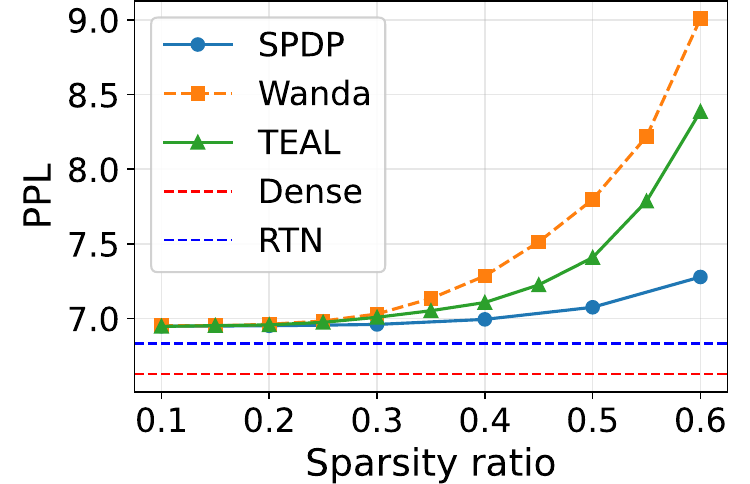}
    \label{fig:downstream_task_result_b}
  }
  % \vspace{-0.5em}
  \caption{Accuracy on downstream tasks and compatibility with quantization. }
  \label{fig:downstream_task_result}
  % \vspace{-1em}
\end{figure*}

\section{Evaluation}

\jhk{To evaluate SPDP, we examine both model quality and system performance. We first describe the experimental setup, including models, pruning procedures, GPU platforms, baselines, and evaluation metrics such as kernel throughput and time per output token (TPOT). We then present kernel-level and end-to-end results, together with profiling analyses that explain them.}
\subsection{Setup}

\jhk{We evaluate the model quality of \frameworktitle{} on the following benchmarks:
WikiText~\cite{wikitext} (perplexity; lower is better),
HumanEval~\cite{Chen2021EvaluatingLL} (\texttt{pass@1}) for code generation,
GSM8K~\cite{Cobbe2021TrainingVT} (exact-match accuracy) for math reasoning,
CoQA~\cite{reddy-etal-2019-coqa} (token-level F1) for conversational QA,
and MMLU~\cite{hendrycks2021measuring} (multiple-choice accuracy) for general knowledge.
We compare against TEAL as a DP baseline and Wanda as a SP baseline.
Unless stated otherwise, all results are reported on \texttt{Llama-2-7B-hf}.
Since \texttt{Llama-2-7B-hf}~\cite{touvron2023llama}
has limited code-generation capability, we evaluate all methods
on HumanEval using \texttt{Qwen3-32B}~\cite{yang2025qwen3technicalreport}.
While evaluating quantization compatibility, we adopt 4-bit round-to-nearest (RTN)
weight-only quantization and follow a fixed pipeline of
SP $\rightarrow$ quantization $\rightarrow$ DP;
DP is enabled at inference time on top of the (pruned and quantized) weights,
and our current pipeline does not support other orderings.}

We evaluate the performance of \frameworktitle{} at two levels: (1) the \frameworktitle{}-\texttt{spMspV} kernel level and (2) the end-to-end framework level. Experiments are conducted on three NVIDIA GPUs: (1) A10G, (2) L4, and (3) L40S. For kernel-level analysis, we use \emph{NVIDIA Nsight Compute}~\cite{ncu} to collect detailed hardware metrics, including instruction throughput, memory efficiency, and shared-memory conflicts. For end-to-end evaluation, each inference is executed 100 times after 10 warm-up iterations, and the mean wall-clock latency is reported.
% We evaluate the performance of \frameworktitle{} at two levels: (1) the \frameworktitle{}-\texttt{spMspV} kernel level and (2) the end-to-end framework level. Experiments are conducted on three NVIDIA GPUs: (1) A10G (24~GB), (2) L4 (24~GB), and (3) L40S (48~GB). For kernel-level analysis, we use \emph{NVIDIA Nsight Compute}~\cite{ncu} to collect detailed hardware metrics, including instruction throughput, memory efficiency, and shared-memory conflicts. For end-to-end evaluation, each inference is executed 100 times after 10 warm-up iterations, and the mean wall-clock latency is reported.
We compare \frameworktitle{} against state-of-the-art sparse inference frameworks, including \texttt{SpInfer}, integrated into the Hugging Face Transformers library~\cite{hf_transformers}. These systems primarily accelerate the decoding phase. We therefore report \emph{Time Per Output Token} (TPOT) as our main latency metric, isolating GPU-side latency using \emph{NVIDIA Nsight Systems}~\cite{nsys} to exclude CPU overhead. 

For the model used during evaluation, we use \texttt{Llama-2-7B-hf} with Wanda~\cite{wanda} as the unstructured SP method and TEAL~\cite{teal} as DP method. The SP method is calibrated \jhk{on} 128 samples of C4 dataset~\cite{c4}, while the DP method is calibrated \jhk{on} 300 samples of Alpaca dataset~\cite{alpaca} to build a layer-wise empirical cumulative distribution function (ECDF) of activation magnitudes. Given a target dynamic sparsity ratio, we set a layer-specific threshold by taking the corresponding ECDF quantile, following TEAL’s uniform sparsity setting. The calibration is performed offline once per model; therefore, the threshold selection process does not introduce any runtime overhead during inference. 
Note that although our experiments are performed on NVIDIA GPUs, the design principles of \frameworktitle{} generalize to other accelerators such as Intel CPUs and NPUs with custom matrix-vector units, providing insights for future sparsity-aware kernel co-design across heterogeneous hardware.

% Although evaluated on NVIDIA GPUs, the design principles of \frameworktitle{} extend to Intel CPUs and NPUs with custom matrix-vector units, informing sparsity-aware kernel co-design across heterogeneous hardware.
\subsection{Model Quality under Downstream Tasks}

\textbf{Model quality under downstream tasks. }
Figure~\ref{fig:downstream_task_result_a} summarizes downstream-task accuracy across diverse workloads. At a moderate sparsity level (e.g., 40\%), \frameworktitle{} matches or closely tracks the baselines across tasks. As sparsity increases to a practically relevant regime for acceleration, \frameworktitle{} consistently degrades less than DP-only (TEAL) and SP-only (Wanda), indicating that unified static-dynamic pruning preserves model quality more robustly across task types while enabling higher effective sparsity.

\textbf{Compatibility with Quantization. }
Figure~\ref{fig:downstream_task_result_b} reports perplexity under 4-bit RTN quantization on \texttt{Qwen3-32B}. We observe that the perplexity trend remains stable across sparsity levels, suggesting that \frameworktitle{} is compatible with weight-only quantization and can potentially stack compression gains without a significant additional perplexity increase. Moreover, combining sparsity with 4-bit weights further reduces model footprint and weight-bandwidth demand, creating additional opportunities for decode-phase TPOT speedups when paired with quantized inference kernels.

% \vspace{-0.5em}
\subsection{Kernel Performance Analysis}
\begin{figure}[t]
  \centering
  \includegraphics[width=0.90\linewidth]{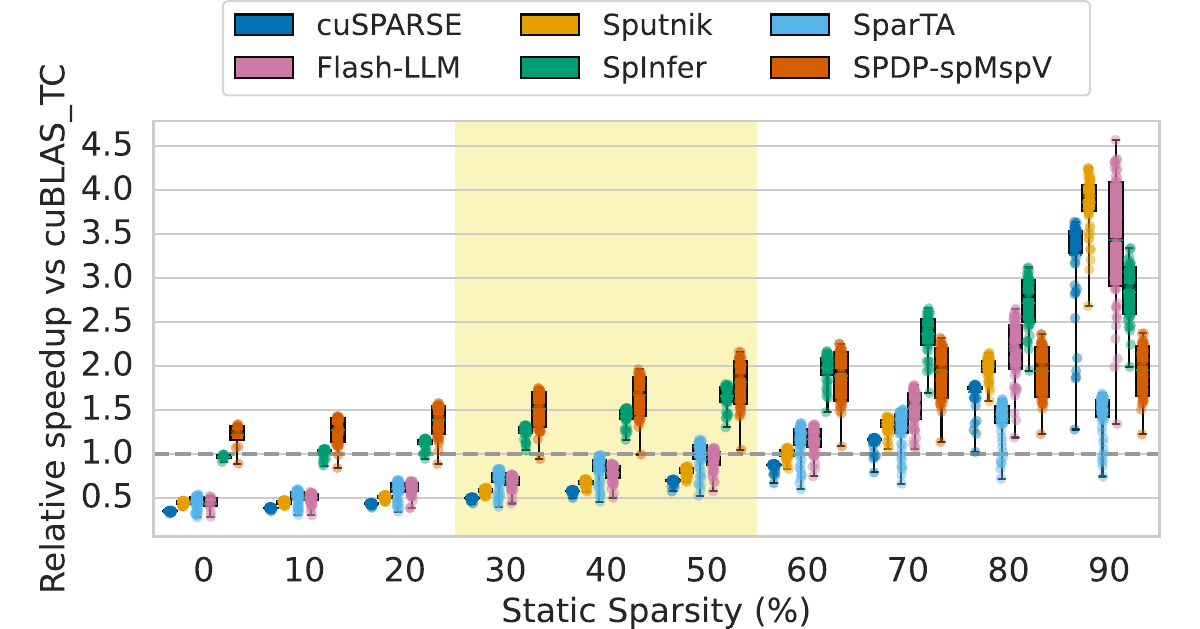}
    % \vspace{-0.3em}
  \caption{Kernel speedup on A10G GPU over cuBLAS\_TC under a static
    sparsity sweep from 0\% to 90\%, with DP fixed at 30\%.
    The shaded
    region marks the moderate-sparsity regime targeted by LLM inference.}
  \label{fig:kernel_sparsity_sweep}
  % \vspace{-1em}
\end{figure}
\textbf{Sparsity-regime characterization.}
\rb{Figure~\ref{fig:kernel_sparsity_sweep} characterizes the decode-kernel performance envelope by sweeping the SP ratio from 0\% to 90\%, while fixing the DP ratio to 30\%. At low static sparsity, the benefit of static compression is limited, although \frameworktitle{} can still exploit the DP by skipping activation-inactive columns.
As static sparsity increases into the moderate regime targeted by LLM pruning, static weight compression and DP jointly reduce memory traffic, while the CUDA-core GEMV path preserves regular register-level accumulation.
At very high static sparsity, some sparse-domain baselines become more competitive because they operate directly on very small nonzero counts, whereas Tiled-CBC still pays bitmap, \texttt{ColInfo}, and tile-level decoding overheads.
% This high-sparsity regime is less relevant for LLM inference, where model quality constraints typically make aggressive unstructured pruning less practical.}
Such high sparsity is less relevant to LLM inference, as model quality
constraints limit aggressive unstructured pruning.}

\rb{We next evaluate practical LLM operating points across weight matrix shapes, where static sparsity lies in the moderate
range and DP provides additional input-dependent column skipping.}

\textbf{LLM projection matrix. }
We compare the \frameworktitle{}-\texttt{spMspV} kernel against six baselines: (1) Tensor-Core-based cuBLAS~\cite{cublas}, which serves as the dense baseline, (2) cuSPARSE~\cite{nvidia_cusparse_2023}, general sparse kernel for diverse sparse application, (3) Sputnik~\cite{sputnik}, (4) SparTA~\cite{sparta}, which targets general deep learning sparse kernel, (5) Flash-LLM~\cite{flash-llm} (6) SpInfer~\cite{spinfer}, a state-of-the-art Tensor-Core-based sparse kernel optimized for moderate sparsity. Following prior work~\cite{flash-llm, spinfer}, we focus on the practical sparsity range of 30–50\% for both unstructured SP and DP, which represents the regime most applicable to large-scale LLMs. 
% Note that other SpMM-based frameworks such as \texttt{FlashLLM}~\cite{flash-llm}, \texttt{Sputnik}~\cite{sputnik}, \texttt{SparTA}~\cite{sparta}, and \texttt{cuSPARSE}~\cite{nvidia_cusparse_2023} are excluded from comparison since their latency exceeds \texttt{cuBLAS} and their compression ratios (CR) exceed~1.0, making fair evaluation impractical.

Note that the kernel is evaluated using the same set of weight matrix shapes adopted in \texttt{SpInfer}~\cite{spinfer}, derived from major LLM architectures including OPT (13B–175B)~\cite{zhang2022optopenpretrainedtransformer}, LLaMA-2 (7B–70B)~\cite{touvron2023llama}, LLaMA-3 (8B, 70B)~\cite{llama3}, Qwen (7B, 14B)~\cite{bai2023qwentechnicalreport}, Qwen2 (7B, 72B)~\cite{yang2024qwen2technicalreport}, and Mixtral-8$\times$7B MoE~\cite{jiang2024mixtralexperts}, Phi-2, 3 (Mini, Medium)~\cite{abdin2024phi3technicalreporthighly}, gpt-oss (20B, 120B)~\cite{openai2025gptoss120bgptoss20bmodel}.
These matrices capture representative dimensions of projection and feed-forward layers in modern Transformer blocks.

\begin{figure}[!t]
  \centering
  \subfloat[Kernel performance on A10G GPU]{
    \includegraphics[width=\linewidth]{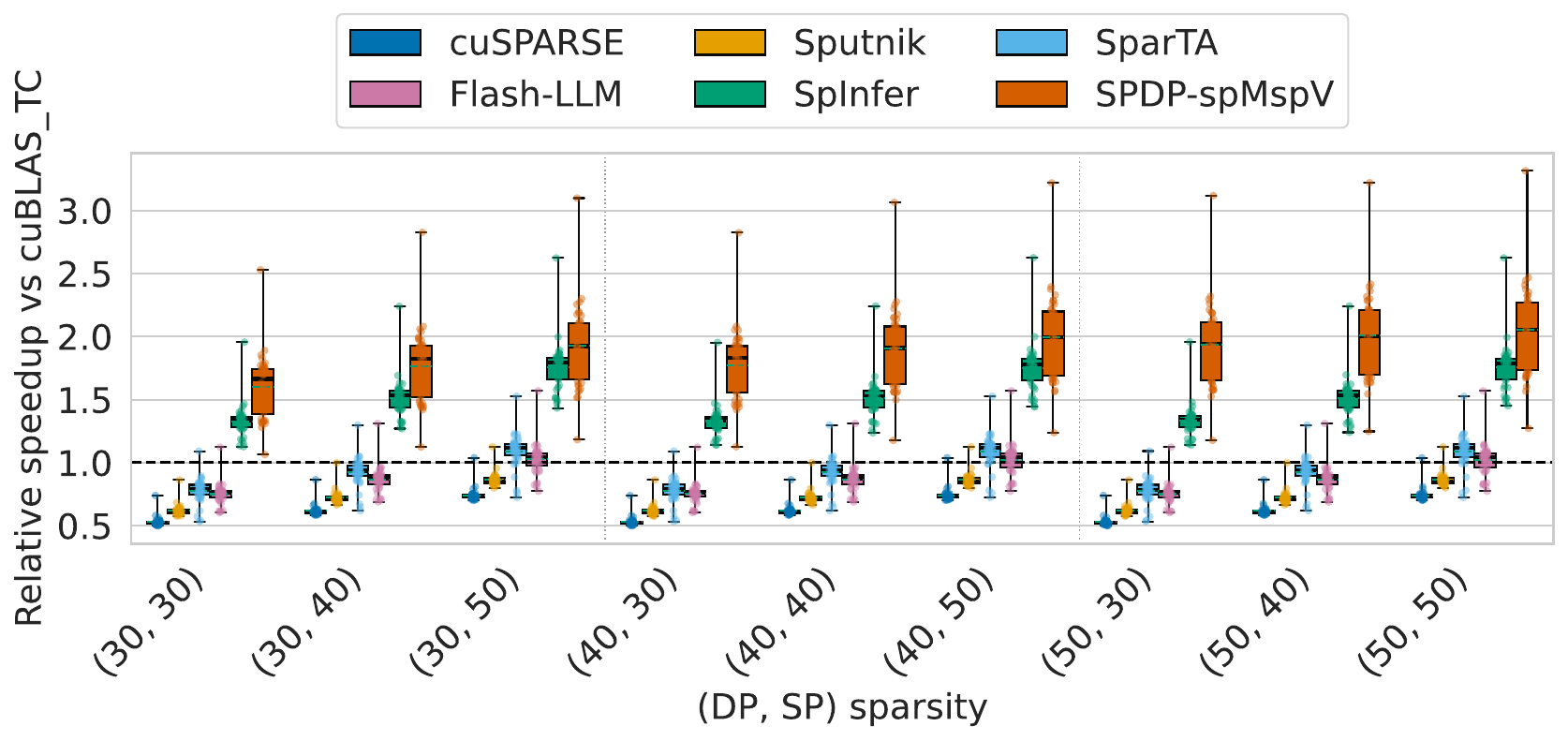}
    \label{fig:kernel_perf_a10g}
  }
  \vspace{0em}
  \subfloat[Kernel performance on a L4 GPU]{
    \includegraphics[width=\linewidth]{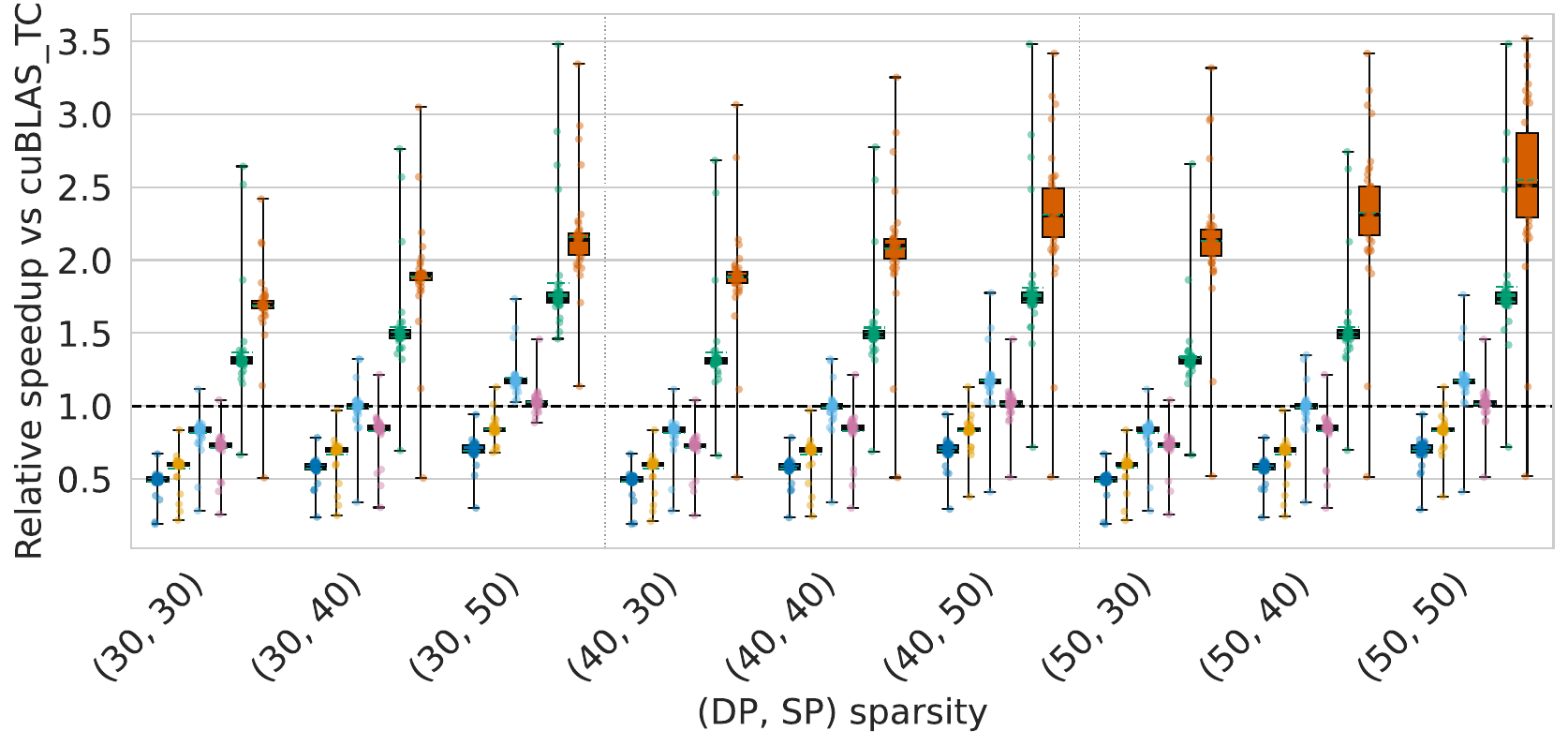}
    \label{fig:kernel_perf_l4}
  }
  % \subfloat[Kernerl performance on L40S GPU]{
  %   \includegraphics[width=0.8\linewidth]{figures_evaluation/Figure_L40S_m8192_k29568_concat.png}
  %   \label{fig:kernel_perf_l40s}
  % }
    \caption{Kernel-level performance comparison of the \frameworktitle{}-\texttt{spMspV} kernel on A10G and L4 GPUs. All speedups are normalized to cuBLAS\_TC.}
  \label{fig:kernel_perf}
  % \vspace{-1.0em}
\end{figure}

Figure~\ref{fig:kernel_perf} reports the kernel throughput (TFLOP/s) normalized to the Tensor-Core-based \texttt{cuBLAS} baseline, indicated by the blue dashed line.
Across all GPU platforms, \frameworktitle{}-\texttt{spMspV} consistently outperforms both dense and sparse baselines.
On the A10G GPU, it achieves an average speedup of 1.88$\times$ over \texttt{cuBLAS} and 1.24$\times$ over \texttt{SpInfer}, with a maximum of 3.32$\times$ and 1.70$\times$, respectively. Similar trends are observed on the L4 GPU, where average speedups reach 2.11$\times$ and 1.37$\times$, and maximum improvements reach 3.52$\times$ and 2.52$\times$ over \texttt{cuBLAS} and \texttt{SpInfer}, respectively. 
The L40S results, used for end-to-end evaluation (Section~\ref{subsec:e2e_llm_eval}), show the same relative ordering, confirming the consistency of the kernel’s scalability.

These results highlight the effectiveness of \frameworktitle{}’s format–kernel co-design.
The \texttt{spMspV} kernel efficiently exploits both static and dynamic sparsity through the \emph{Tiled-CBC} layout, achieving high compute intensity and minimizing index overhead.
By aligning memory access with column-wise activation sparsity, it maintains coalesced data movement and reduces metadata traffic, which are dominant factors in GPU-bound workloads.

Notably, DP provides an additional efficiency gain by skipping inactive neurons at runtime, improving both computational density and memory locality.
This adaptivity enables \frameworktitle{}-\texttt{spMspV} to outperform \texttt{SpInfer}, which is limited to static sparsity.
Consequently, \frameworktitle{} bridges the gap between static compression and activation-dependent sparsity, achieving high throughput and scalability across diverse GPU architectures.

\subsection{Kernel Hardware Analysis}

\begin{figure}
  \centering
    \includegraphics[width=0.95\linewidth]{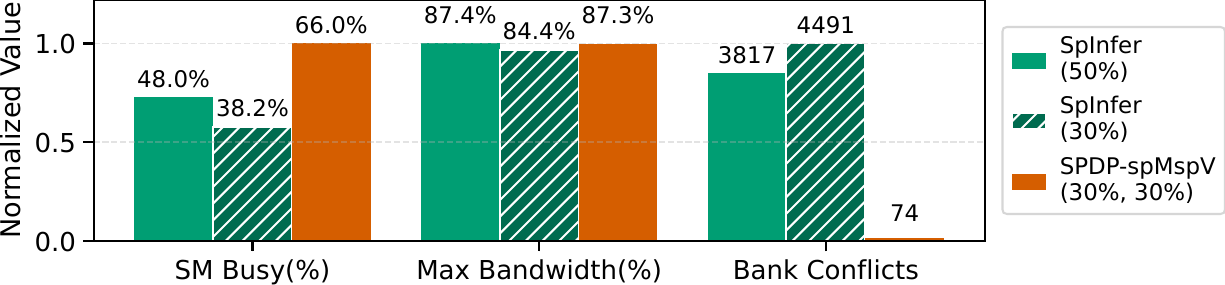}
    % \vspace{-0.2em}
\caption{Kernel profiling results on the A10G GPU. \frameworktitle{}-\texttt{spMspV} uses (DP, SP) = (30\%, 30\%), compared with SpInfer (30\%, same SP) and SpInfer (50\%, same total sparsity). All metrics are normalized to the highest value.}
  \label{fig:kernel_analysis}
  % \vspace{-0.5em}
\end{figure}

% Summary
Figure~\ref{fig:kernel_analysis} profiles GPU utilization (\emph{SM busy}, \emph{Max bandwidth}, and \emph{Bank conflicts}) for \frameworktitle{}-\texttt{spMspV} with $(\mathrm{DP},\mathrm{SP})=(30\%,30\%)$ and \texttt{SpInfer} with $\mathrm{SP}=30\%$ and $50\%$.
% Figure~\ref{fig:kernel_analysis} profiles GPU utilization (\emph{SM busy}, \emph{Max bandwidth}, and \emph{Bank conflicts}) for \frameworktitle{}-\texttt{spMspV} with $(\mathrm{DP},\mathrm{SP})=(30\%,30\%)$ and \texttt{SpInfer} with $\mathrm{SP}=30\%$ and $50\%$ (\texttt{SpInfer(30\%)}/\texttt{SpInfer(50\%)}).
\emph{SM busy} is the fraction of cycles where any SM unit is active (preferred over Instruction Per Cycle (IPC) since Tensor Core \texttt{mma} instructions naturally yield lower IPC), \emph{Max bandwidth} is achieved DRAM bandwidth normalized to peak, and \emph{Bank conflicts} captures serialized shared-memory transactions.

\frameworktitle{}-\texttt{spMspV} attains the highest \emph{SM busy}, indicating that combining SP and DP better sustains compute--memory overlap by skipping redundant activation reads; \texttt{SpInfer(30\%)} is lower without DP, and \texttt{SpInfer(50\%)} improves only modestly, highlighting the benefit of runtime adaptivity. All three kernels achieve comparable Max bandwidth, with 
\frameworktitle{}-\texttt{spMspV} slightly lower than both 
\texttt{SpInfer(50\%)} and \texttt{SpInfer(30\%)}. 
This phenomenon indicates that decompression and DP introduce negligible 
memory-stall overhead while preserving effective computation–memory overlap. Finally, \frameworktitle{}-\texttt{spMspV} achieves the lowest \emph{Bank conflicts} thanks to the Tiled-CBC column-aligned tiling, which yields more regular shared-memory accesses than \texttt{SpInfer}'s SMBD under mixed sparsity. Overall, these hardware-level efficiencies explain the throughput advantage of \frameworktitle{} over dense and static-sparse baselines.

\subsection{End-to-end LLM Inference Evaluation}\label{subsec:e2e_llm_eval}
% \begin{figure}
%   \centering
%   \subfloat[A10G]{
%     \includegraphics[height=2.23cm]{figures_evaluation/tpot_vs_ppl_A10G.pdf}
%     \label{fig:tpot_ppl_a10g}
%   }
%   \hfill
%   \subfloat[L4]{
%     \includegraphics[height=2.23cm]{figures_evaluation/tpot_vs_ppl_L4.pdf}
%     \label{fig:tpot_ppl_l4}
%   }
%   \hfill
%   \subfloat[L40S]{
%     \includegraphics[height=2.23cm]{figures_evaluation/tpot_vs_ppl_L40S.pdf}
%     \label{fig:tpot_ppl_l40s}
%   }
%     \caption{Perplexity vs. TPOT on A10G, L4, and L40S GPUs. SPDP attains lower TPOT with a 25\% sparsity ratio difference, suggesting improved memory efficiency.}
%   \label{fig:ppl_vs_tpot}
%   % \vspace{-0.5em}
% \end{figure}
\begin{figure}
  \centering
    \includegraphics[width=\linewidth]{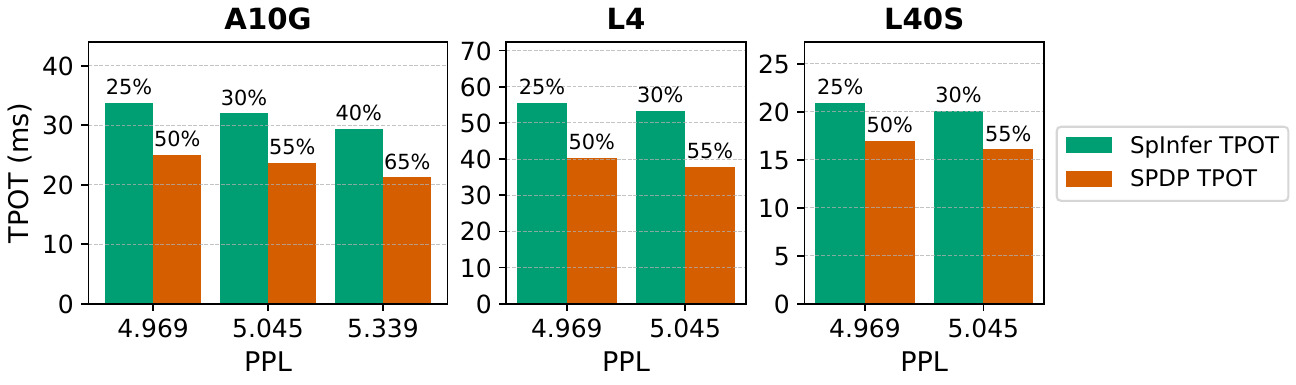}
    % \vspace{-0.2em}
\caption{Perplexity vs. TPOT on A10G, L4, and L40S GPUs. SPDP attains lower TPOT with a 25\% sparsity ratio difference, suggesting improved memory efficiency.}
  \label{fig:ppl_vs_tpot}
  % \vspace{-0.5em}
\end{figure}

% \begin{figure*}[htbp]
%   \centering
%   \captionsetup[subfloat]{captionskip=0pt}
%   % ----- SPDP (Top Row) -----
%   \subfloat[A10G (SPDP)]{
%     \includegraphics[width=0.32\linewidth]{figures_evaluation/tpot_ppl_vs_sparsity_from_csv_A10G.pdf}
%   }
%   \hfill
%   \subfloat[L4 (SPDP)]{
%     \includegraphics[width=0.32\linewidth]{figures_evaluation/tpot_ppl_vs_sparsity_from_csv_L4.pdf}
%   }
%   \hfill
%   \subfloat[L40S (SPDP)]{
%     \includegraphics[width=0.32\linewidth]{figures_evaluation/tpot_ppl_vs_sparsity_from_csv_L40S.pdf}
%   }
  
%   % \vspace{-0.7em}

%   % ----- SpInfer (Bottom Row) -----
%   \subfloat[A10G (SpInfer)]{
%     \includegraphics[width=0.32\linewidth]{figures_evaluation/tpot_ppl_vs_sparsity_spinfer_from_csv_A10G.pdf}
%   }
%   \hfill
%   \subfloat[L4 (SpInfer)]{
%     \includegraphics[width=0.32\linewidth]{figures_evaluation/tpot_ppl_vs_sparsity_spinfer_from_csv_L4.pdf}
%   }
%   \hfill
%   \subfloat[L40S (SpInfer)]{
%     \includegraphics[width=0.31\linewidth]{figures_evaluation/tpot_ppl_vs_sparsity_spinfer_from_csv_L40S.pdf}
%   }

%     \caption{Sparsity ratio vs. TPOT speedup across different GPU architectures. SPDP reduces TPOT while maintaining reasonable perplexity, whereas SpInfer diverges due to the inherent limitation of SP. Beyond 65\% sparsity, Wanda’s perplexity diverges drastically, rendering SpInfer impractical in this regime.}
%   \label{fig:s_vs_tpot}
%   % \vspace{-0.5em}
% \end{figure*}

\begin{figure*}[htbp]
  \includegraphics[width=0.90\linewidth]{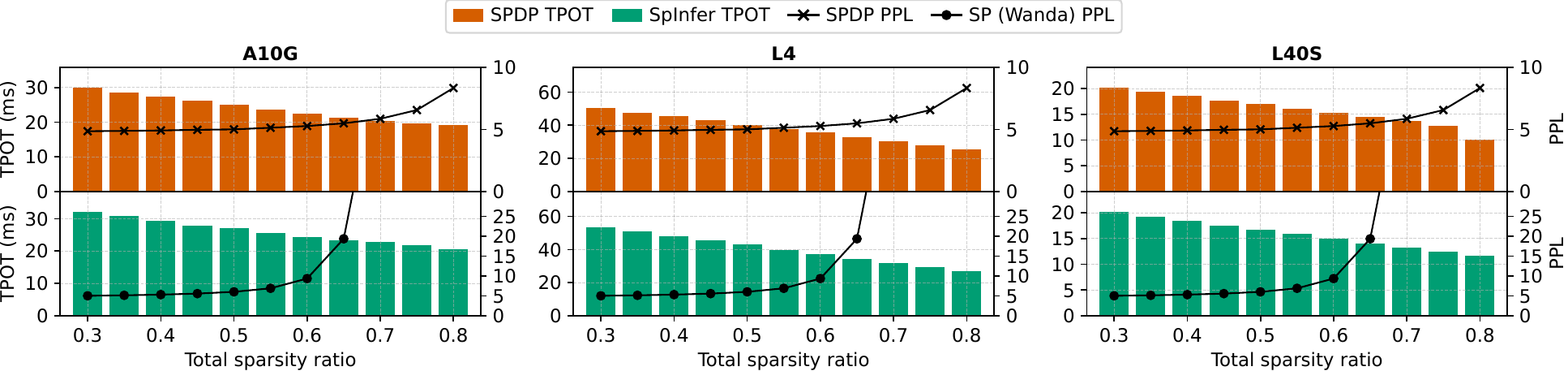}

    \caption{Sparsity ratio vs. TPOT speedup across different GPU architectures. SPDP reduces TPOT while maintaining reasonable perplexity, whereas SpInfer diverges due to the inherent limitation of SP. Beyond 65\% sparsity, Wanda’s perplexity diverges drastically, rendering SpInfer impractical in this regime.}
  \label{fig:s_vs_tpot}
  % \vspace{-0.5em}
\end{figure*}

\textbf{Sparsity Ratio vs. TPOT Speedup.}
% Figure~\ref{fig:s_vs_tpot} shows the end-to-end \emph{Time Per Output Token (TPOT)} speedup of the \texttt{Llama-2-7B-hf} model across varying sparsity ratios and GPUs.
% For each configuration, \frameworktitle{} determines the optimal combination of static and DP that minimizes perplexity.
% For instance, to achieve an overall sparsity of 65\%, combining 30\% unstructured SP (\emph{Wanda}) with 50\% DP (\emph{TEAL}) yields the best model quality.
% \emph{Wanda} serves as the unstructured SP method for both \frameworktitle{} and \texttt{SpInfer}, ensuring a consistent baseline.
\jhk{Figure~\ref{fig:s_vs_tpot} shows the end-to-end \emph{Time Per Output Token (TPOT)} speedup of the \texttt{Llama-2-7B-hf} model across varying sparsity ratios and GPUs.
For each configuration, \frameworktitle{} determines the optimal combination of SP and DP that minimizes perplexity.
For instance, the best configuration at 65\% total sparsity uses 30\% SP and 50\% DP.}
% Both \frameworktitle{} and \texttt{SpInfer} use \emph{Wanda} for SP.

As sparsity increases, both frameworks show diminishing TPOT speedups accompanied by rising perplexity.
Perplexity escalates sharply beyond 70\% static sparsity, where essential model weights are excessively pruned.
Unlike kernel-level evaluations, this experiment measures end-to-end decode latency—including attention, normalization, and other operations.
\rb{Many of these components are not sparsity-sensitive, so overall acceleration follows an Amdahl-style upper bound}: even large speedups in sparse kernels translate to modest end-to-end gains.
\rb{Decoding workloads are also primarily memory-bound}; latency is dominated by data movement rather than computation.
Thus, increasing sparsity improves arithmetic efficiency but only moderately reduces latency, as performance saturates under limited memory bandwidth.
Despite these constraints, \frameworktitle{} achieves higher TPOT speedup (average $\times$1.34) than \texttt{SpInfer} at comparable and moderate sparsity, while maintaining lower perplexity (under 5.4).
This demonstrates that integrating DP enables finer-grained adaptivity to input-dependent redundancy, yielding a better speed–quality trade-off than static sparsity alone.

\textbf{Perplexity vs. TPOT Speedup.}
While Figure~\ref{fig:s_vs_tpot} captures the aggregate impact of sparsity on performance and quality, it does not isolate the effect of DP under matched model quality.
To ensure a fair comparison, we analyze TPOT speedup as a function of perplexity at equivalent accuracy levels.

As shown in Figure~\ref{fig:ppl_vs_tpot}, at the same perplexity, \frameworktitle{} attains substantially higher overall sparsity—up to 25\% more—than \texttt{SpInfer} (Wanda-based SP), while achieving comparable or lower TPOT. This indicates that DP effectively exploits activation-dependent sparsity without compromising model fidelity.
Although additional sparsity does not drastically reduce latency due to the memory-bound nature of decoding, it meaningfully decreases data movement, thereby improving \emph{energy efficiency}. By transferring and processing fewer bytes per token, \frameworktitle{} delivers higher performance-per-watt and better scalability in large-scale serving scenarios.

Overall, these results demonstrate that the synergy of static and dynamic pruning in \frameworktitle{} not only reduces token-generation latency but also enhances energy efficiency and quality preservation under realistic, memory-limited inference conditions.

% \vspace{-0.2em}
\section{Discussion and Future Work}

\textbf{System-level generalization.}
\frameworktitle{} is built around an execution model that combines unstructured SP with threshold-based DP in memory-bound decode. Many DP methods rely on threshold-based selection~\cite{federici2025efficient, cats}, and can reuse SPDP’s column-wise skipping with minor interface changes (e.g., how thresholds or per-token scores are produced). DP variants that materialize explicit masks/bitmaps can also be accommodated by mapping them to the same column-selection interface.

\begin{figure}
    \centering
    \includegraphics[width=0.9\linewidth]{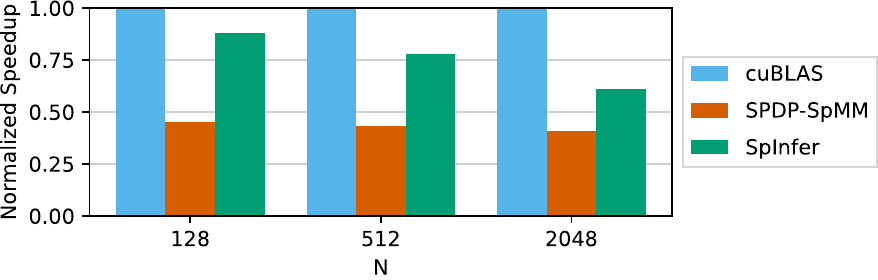}
    % \vspace{-1.0em}
    \caption{Prefill kernel comparison on A10G under SP=30\%. With large N, existing works fail to beat the cuBLAS baseline.}
    \label{fig:prefill}
    % \vspace{-1.2em}
\end{figure}

Prefill typically has a large output width ($N=B\times L$), which increases CI and often shifts the workload toward a compute-bound regime; as a result, LSCD-style sparse prefill does not consistently outperform dense cuBLAS\_TC, and decoding/layout-alignment overheads become more visible, as shown in Figure~\ref{fig:prefill}. Accordingly, SPDP’s prefill path mainly prioritizes \emph{unified-format compatibility} (reusing a single compressed representation across prefill and decode without reformatting) and \emph{memory-footprint reduction}, while the primary latency benefit comes from the bandwidth-bound decode stage. In a prefill--decode disaggregated deployment~\cite{distserve, splitwise}, a practical option is to run dense GEMM for prefill and use SPDP-compressed weights only for decode.

Integrating SPDP into modern LLM serving engines (e.g., continuous batching or fused prefill--decode) is largely orthogonal to attention execution, since SPDP mainly replaces the projection matmul kernels while leaving attention kernels unchanged. 

SPDP kernels and the Tiled-CBC format are compatible with quantization, since \texttt{Values} are stored separately from sparsity metadata. However, quantization and unstructured sparsity do not provide multiplicative compression in practice: low-bit precision reduces only \texttt{Values}, while \texttt{Bitmap}, \texttt{ColInfo}, \texttt{TileOffset}, and padding remain. As a result, metadata can occupy a large fraction of storage at low precision, making such stacking beneficial mainly at very high sparsity. Exploring coarser-grained static sparsity, such as block or semi-structured patterns, to better amortize metadata under quantization remains future work.

% Finally, while Tiled-CBC can store low-bit payloads in principle, under unstructured sparsity the bitmap and metadata overhead becomes relatively larger at low precision, limiting the practical stacking benefit with weight-only quantization. Exploring higher-granularity static sparsity (e.g., block or semi-structured patterns) to better amortize metadata and align with quantization remains future work. 

\begin{figure}[t]
  \centering
  \includegraphics[width=\linewidth]{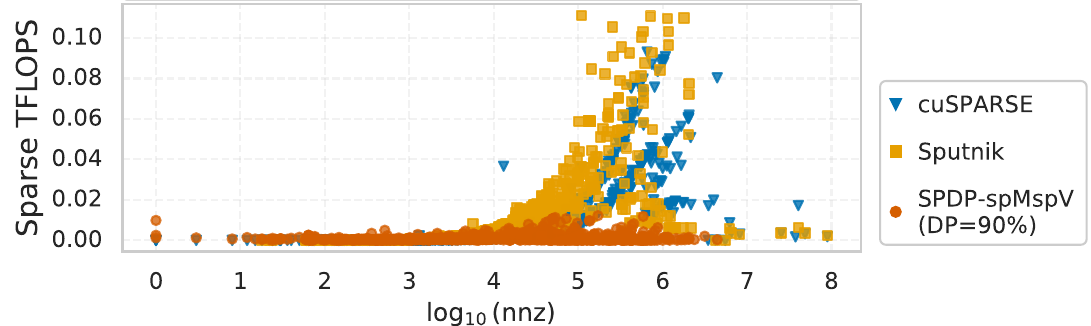}
    % \vspace{-1.0em}
  \caption{SuiteSparse evaluation. SPDP targets moderate-sparsity LLM decode workloads, whereas many SuiteSparse matrices are extremely sparse.}
  \label{fig:suitesparse_perf}
  % \vspace{-1.2em}
\end{figure}

\textbf{Workload scope and sparsity regime.}
\rb{To clarify whether SPDP is a general-purpose sparse linear algebra kernel, we additionally evaluate non-LLM sparse matrices from the SuiteSparse Collection~\cite{suitesparse}. Figure~\ref{fig:suitesparse_perf} compares SPDP-\texttt{spMspV} against cuSPARSE and Sputnik on sparse matrix-vector workloads.}
\rb{SPDP is not consistently faster on these matrices. This is expected: many SuiteSparse matrices operate in an extreme-sparsity regime, with roughly 85\% of evaluated matrices exceeding 95\% sparsity, where CSR/CSC-style sparse-domain execution avoids reconstructing dense tiles and better amortizes index overhead.
In contrast, SPDP targets the moderate-sparsity regime typical of LLM pruning, where model quality constraints prevent extreme sparsity and regular tile-level execution remains beneficial. Thus, SPDP should be viewed as a specialized LLM-decode kernel rather than a replacement for general sparse linear algebra libraries.}

% \rb{SPDP is not consistently faster on these matrices. This is expected: many SuiteSparse matrices operate in an extreme-sparsity regime, with roughly 85\% of the evaluated matrices exceeding 95\% sparsity, where CSR/CSC-style sparse-domain execution can avoid reconstructing dense tiles and therefore better amortizes index overhead.
% In contrast, SPDP is designed for the moderate-sparsity regime typical of LLM pruning, where model quality constraints prevent extreme sparsity and where regular tile-level execution is still beneficial. Thus, SPDP should be viewed as a specialized LLM-decode kernel rather than a replacement for general sparse linear algebra libraries.}

\section{Conclusion}

We presented \frameworktitle{}, a unified sparse-inference framework that exploits unstructured static pruning (SP) and input-dependent dynamic pruning (DP). At the core of \frameworktitle{} is the Tiled-CBC format, which enables runtime sparse-fragment access for decode while remaining reusable for prefill. Built on this representation, \frameworktitle{} provides a CUDA-core \texttt{spMspV} kernel for decode and a Tensor-Core \texttt{SpMM} kernel for prefill, allowing a single compressed format to support both phases of LLM inference. Our evaluations show that \frameworktitle{} improves kernel throughput, TPOT, and power efficiency over prior sparse baselines while maintaining model quality. Future work includes extending \frameworktitle{} to batched dynamic pruning, production serving integration, and multi-GPU sparsity-aware scheduling.

\begin{acks}

 This work was supported by the National Research Foundation of Korea (NRF; RS-2025-00560762 and RS-2024-00414981), the Institute of Information \& Communications Technology Planning \& Evaluation (IITP; RS-2024-00454666, RS-2025-25442338, RS-2021-II211343, RS-2026-25551943, RS-2026-25518317, and RS-2024-00397085), the Ministry of Health and Welfare under the ARPA-H BAYS Project (RS-2025-25455095), the Ministry of Science and ICT under the Advanced GPU Utilization Support Program, and Samsung Electronics Co., Ltd. (IO251215-14535-01). The authors thank Samsung MAX Lab for providing research infrastructure. J. Do is affiliated with the Automation and Systems Research Institute (ASRI), Seoul National University (SNU).
\end{acks}

%\clearpage
\balance
\bibliographystyle{ACM-Reference-Format}
\bibliography{sample}

\end{document}